%% file: aanda.tex
\documentclass{aa}  

\usepackage{graphicx}
\usepackage{subcaption}
\usepackage{placeins}
\usepackage{lscape}
\usepackage[colorlinks=true,linkcolor=blue,citecolor=blue]{hyperref}
\usepackage{txfonts}

\begin{document}

   \title{Multi-messenger observations of binary neutron star mergers: synergies between the next  generation gravitational wave interferometers and wide-field, high-multiplex spectroscopic facilities
   }
   \subtitle{}

   \author{S. Bisero\inst{1}
          \and
          S. D. Vergani\inst{1}
          \and
          E. Loffredo\inst{2,3,4}
          \and
          M. Branchesi \inst{3,4}
          \and
          N. Hazra \inst{5,3}
          \and
          U. Dupletsa \inst{3,4,6} 
          \and
          R. I. Anderson \inst{7}
          }

   \institute{LUX, Observatoire de Paris, Université PSL, Sorbonne Université, CNRS, 92190 Meudon, France
        \and
        INAF – Osservatorio Astronomico d’Abruzzo, 64100 Teramo, Italy
        \and
        Gran Sasso Science Institute (GSSI), I-67100 L'Aquila, Italy
        \and
        INFN, Laboratori Nazionali Del Gran Sasso,  I-67100 Assergi, Italy
        \and 
        National Centre for Nuclear Research, Pasteura 7, PL-02-093 Warsaw, Poland
        \and
        Institute of High Energy Physics - Austrian Academy of Sciences, 1010 Vienna, Austria
        \and
        Institute of Physics, Laboratory of Astrophysics, École Polytechnique Fédérale de Lausanne (EPFL), Observatoire de Sauverny, 1290 Versoix, Switzerland 
             }

   \date{
   }

\abstract
{Third-generation gravitational wave (GW) observatories such as the Einstein Telescope (ET) and Cosmic Explorer (CE), will access a large volume of the Universe, detecting hundreds of thousands of binary neutron star (BNS) mergers, reaching distances beyond $z \sim 3$. The unique information revealed by joint GW and electromagnetic (EM) detections can be fully exploited only with a dedicated observing strategy and suitably adapted EM facilities.}
{In this work, we explore the impact of Integral Field and Multi-Object Spectroscopy (IFS and MOS) with the Wide-field Spectroscopic Telescope (WST) on next-generation GW multi-messenger observations. }
{We consider populations of BNS mergers assuming different equations of state and mass distributions, and compute the corresponding GW signals for both the ET operating alone and in a network with CE. Kilonova (KN) light curves are assigned using either AT2017gfo-like models or numerical-relativity-informed ones; gamma-ray burst (GRB) afterglow emission is also included. 
We consider two main observing strategies: one in synergy with wide-field photometric surveys, and a second based on a galaxy-targeted approach, exploiting WST’s high multiplexing capabilities. We estimate the number of galaxies within the GW error volume and discuss potential observational challenges and corresponding mitigation strategies.}
{We find that KNe can be detected with WST up to $ z \sim 0.4 $ and magnitudes $ m_{\mathrm{AB}} \sim 25 $, while GRB afterglows may be observable at higher redshifts, at $ z > 1 $, for viewing angles $ \Theta_{\mathrm{view}} \lesssim 15^\circ $. 
We show that Target of Opportunity observations aimed at KN detection can be optimally scheduled 12--24 hours after the merger. For GRB afterglows, particularly in cases of poor localisation in the first minutes after the high-energy detection, WST IFS can play a key role in counterpart identification and position refinement.
Our results show that mini-IFUs (fibre bundles) and galaxy catalogues complete in redshift up to $ z \leq 0.5 $ will be essential for the efficient identification of EM counterparts. 
We find that the WST contribution is valuable also for events with good sky localisation, because the uncertainties in luminosity distance —even at low redshift — can
result in large error volumes containing numerous galaxies to target -  up to thousands at $z<0.1$ and tens of thousands at $z < 0.2$ for GW detections of ET in a network with CE.
Finally, we underline that events at redshifts $z < 0.3$ and with sky localisations better than $10$\,deg$^2$ will be {\it golden events} for WST, because
it will be possible for WST to cover all the galaxies in the error volume with a limited number of exposures. We estimate them
to be from $\sim10$ (ET-alone configuration) to hundreds per
year (ET+CE configuration). }
{The detection and characterisation of the EM counterparts of BNS detected in the extended volume of the universe explored by next-generation interferometers will be challenging. Observational strategies, selection criteria, and useful telescopes / instruments should be developed in advance. We showed that spectroscopic facilities with large field-of-view, high-
sensitivity, and high-multiplexing modes capable of handling the necessary depths, large sky regions, and vast numbers of objects to be targeted, are powerful instruments to exploit successfully the new multi-messenger science opportunities that will be opened by next-generation GW interferometers.} 


\titlerunning{BNS MM observations with next-generation GW detectors and wide-field, high-multiplex spectroscopic facilities}

\authorrunning{S. Bisero et al.}

\maketitle
%
\section{Introduction}
The extraordinary joint detection of gravitational waves (GWs) and light from a binary neutron star system merger (BNS) on the 17$^{th}$ of August 2017 \citep{abbott17} opened the era of GW multi-messenger (MM) astrophysics. That day, GWs from a BNS merger at a luminosity distance of $\sim$40 Mpc were detected by Advanced Ligo and Virgo detectors, that localised the source within a 28 deg$^{2}$ credible region \citep{LVK170817}. After $\sim$ 1.7 seconds a short gamma-ray burst (SGRB) was detected by \textit{Fermi} and INTEGRAL satellites \citep{goldstein17,savchenko17, LVKGRB}. This sparked an extensive multi-wavelength EM follow-up campaign: a few hours later telescopes from all over the world were tiling the GW sky localisation, including galaxy targeted searches, and found a faint fast evolving optical transient in the galaxy NGC4993 \citep{coulter17}, whose redshift was consistent with the GW source luminosity distance. Subsequent monitoring of the source lightcurve and spectrum, especially the spectroscopic observations taken with VLT/X-shooter from $\sim$ 1.4 days after the GW detection, allowed the classification of the transient as a kilonova \citep[KN;][]{pian17,smartt17}, i.e. thermal radiation powered by the radioactive decay of r-process heavy elements produced in the neutron rich ejecta environment, favored by the merger of the two neutron stars (NSs). Also, from $\sim$ 10 days post-merger on, a broad-band non thermal source was detected at the very same position of the KN and, thanks to VLBI proper motion measurements, and to radio and X-ray campaigns, it is now known to be the off-axis afterglow produced by the structured jet of SGRB170817A  \citep{mooley18,davanzo18,margutti18,ghirlanda19,mooley22}.
This plethora of observations confirmed the theoretical hypothesis that at least part of SGRBs originate in BNS mergers, and raised many open questions concerning, among others, the nature of heavy elements produced in these violent explosions, the population of KN in all its variety, and the link between BNS and SGRBs. Moreover, the study of joint GW-EM detections has implications in various fields: from cosmology, allowing to give independent estimates of the Hubble constant \citep{abbottH0}, to the study of GRB jets structure \citep[e.g.,][]{troja18,lamb19, ghirlanda19}.
Unfortunately, no analogous opportunity to observe simultaneously these two cosmic messengers occurred so far. Before the start of the fourth observing run (O4) of the LIGO/Virgo/KAGRA (LVK) detectors, the expected detection rate of BNS was ${10}_{-10}^{+52}$ yr$^{-1}$ \citep{2020BNSrates}. Unluckily, no BNS detection occurred during the O4 run so far. 

This shows that predicted and effective detection rates of BNS with current GW interferometers are still very low. 
However, with next generation GW observatories, such as the Einstein Telescope (ET, \citealt{ET}) and Cosmic Explorer (CE, \citealt{CE}), they will dramatically increase: with ET alone it will be possible to detect up to $\sim$10$^{5}$ BNS per year well beyond the Local Universe \citep{branchesi23, ET_bluebook}. The majority of sky localisation regions of such events will be large and the EM counterparts will be faint. Facilities that combine high sensitivity with wide fields of view, such as the Vera Rubin Observatory for photometry, will be indispensable for the research of the EM counterparts.
Spectroscopic observations,  fundamental to identify and characterise counterpart candidates, and therefore to discriminate among them, will likely be the bottle-neck of GW MM science. 

In this work we explore the use of integral field and multi-object spectroscopy (IFS and MOS) for the search and identification of EM counterparts of GW detections from next generation GW interferometers such as ET.
IFS and MOS provide a significant advantage by enabling the simultaneous acquisition of multiple spectra.
An IFS captures a spectrum at each spatial pixel within a two-dimensional field, enabling simultaneous spatial and spectral coverage over a compact region of the sky. In contrast, MOS acquires spectra from multiple individual targets across a wide field of view in a single exposure, typically using optical fibres.

To date, the best project of such facilites, is the Wide-field Spectroscopic Telescope (WST, \citealt{bacon24, WST}), a concept study to be realised in the southern hemisphere in the 2040s, overlapping with ET operations. WST will be equipped both with a panoramic IFS, having a FoV nine times larger than that of MUSE, 30\,000 low- resolution fibres and 2\,000 high- resolution fibres (MOS).
It is envisioned as a dedicated 12-meter-class spectroscopic facility, having time-domain and multi-messenger among the structuring science cases. 

We use simulations of BNS detections datasets and corresponding KN and GRB afterglow emission (described in Sect. \ref{sim}) to explore the observations with WST IFS and MOS (Sect. \ref{wstsim}). We study the detectability and characterisations with WST of such EM counterparts, and we investigate how the results depend on the observable and intrinsic properties of the BNS population (Sect. \ref{results}). 
Finally, we discuss different observing strategies, examine potential observational challenges, and explore some selection criteria (Sect.~\ref{obs_strategy}).
We summarise our results in Section \ref{conclusion}.
Throughout this paper, we assume a flat $\Lambda$CDM cosmology with $\mathrm{H_{0}}$ = 67.66km\,s$^{-1}$\,Mpc$^{-1}$, $\Omega_{\mathrm{M}}$ = 0.3 \citep{planck18}.

\section{EM counterparts of BNS mergers detected by ET and CE}\label{sim}

\renewcommand{\arraystretch}{1.5}

\begin{table}[ht]
\small
\begin{center}
\begin{tabular}{|p{0.3cm}|p{1.9cm}|p{0.5cm}|p{1.9cm}|p{1cm}|p{0.7cm}|}
    \hline
     Set & Network \hspace{0.8cm} configuration &  Obs [yrs]& Counterparts  & $\Omega_{90}$ [$\deg^{2}$] & $z$ \\
    \hline
    1 & ET $\Delta$ and $2L$ & 1 & gfo-like KN & $<$ 100 & $<$ 1.2\\
    \hline
    2 & ET $\Delta$ and $2L$ & 10 & theoretical KN (+ GRB) & $<$ 100 & $<$ 1\\
    \hline
    3 & ET $\Delta$ and $2L$ + 1CE & 10 & theoretical KN (+ GRB) & $<$ 40 & $<$ 1\\
    \hline
\end{tabular}
\end{center}
\caption{Overview of the characteristics of the different simulated dataset used for our study.}
\label{sample}
\end{table}

To model the KN and SGRB afterglow signals associated with detections of BNS mergers by ET and CE, we consider three different datasets presented in Section 4.3 of \cite{branchesi23} (Set 1) and in \cite{loffredo25} (Set 2 and 3). 
The KN associated with the GW-detected mergers in Set 1 are modelled based on AT2017gfo, the only KN detected in association with GWs so far. 
Instead, KN lightcurves associated with BNS mergers in Set 2 and 3 are computed based on the results of numerical-relativity simulations targeted to GW170817 and GW190425 \citep{190425}, the second BNS merger detected by LVK. Additionally, Sets 2 and 3 include SGRB afterglow emission.
Despite being more model-dependent, Sets 2 and 3 account for the expected diversity in KN signals from a population of BNS mergers, which can realistically differ significantly from GW170817. 
Below, we briefly report additional information on the three datasets, and refer readers to \cite{branchesi23} and \cite{loffredo25} for more details.

\begin{figure*}
\begin{center}
    \includegraphics[width = 0.75\linewidth]{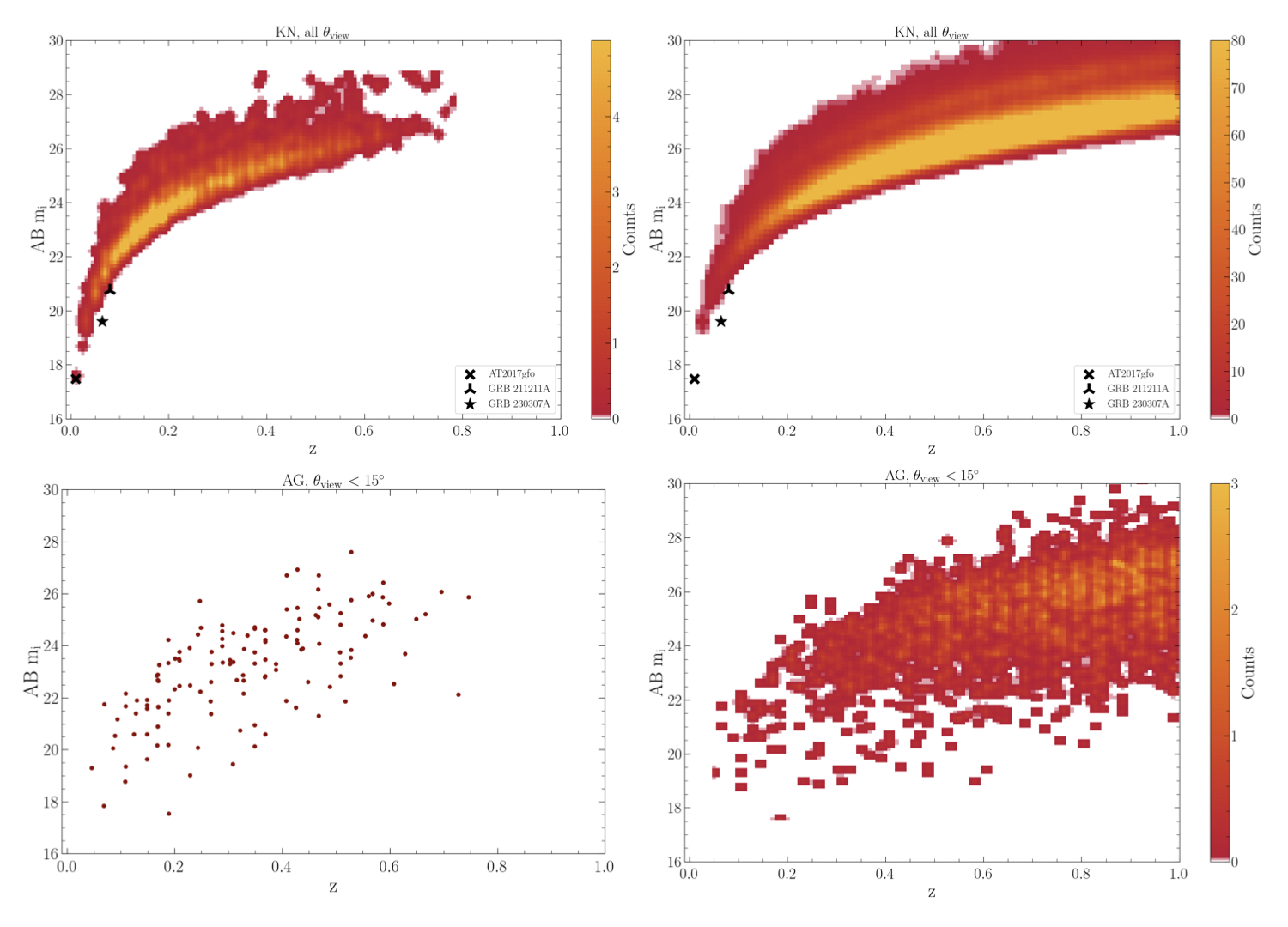}
    \caption{AB magnitude in Rubin $i$ filter at $\sim$12 hours post-merger as a function of redshift for KNe and on-axis ($\theta_{\mathrm{view}}<15^{\circ}$) GRB afterglows following BNS detected by ET as a standalone observatory (left panels) and in a network with CE (right panels). The few bright KNe ($m_{\mathrm{AB}}<21$) in the upper-right panel are not visible due to their low density relative to the bulk of the population, which dominates the color scale. The cross, upward-pointing triangle, and star represent respectively AT2017gfo, and the optical counterparts of GRB~211211A \citep{rastinejad22} and GRB~230307A \citep{levan24}, two GRBs followed by a KN, at $\sim$12 hours after the GRB prompt detection. A Gaussian filter was applied to the data in the top panels and bottom-right one for visualization purposes.}
    \label{mag_vs_z}
    \end{center}
\end{figure*}

\subsection{BNS, GW detection and KN population simulations : Set~1}

The starting BNS merger population is documented in \cite{santoliquido21} and corresponds to a local merger rate $\mathcal{R}_{\mathrm{BNS}}$ = $365 \mathrm{Gpc^{-3}yr^{-1}}$ (common envelope parameter $\alpha = 3$), which is compatible with the 90$\%$ credible interval $\mathcal{R}_{\mathrm{BNS}}$ $\in [10 - 1700] \mathrm{Gpc^{-3}yr^{-1}}$ reported in GWTC-3 \citep{abbott23_GWTC3}. From this population, \cite{ronchini22} and \cite{branchesi23} built a catalogue of one year of BNS mergers, corresponding to $9 \times 10^{5}$ BNS mergers observable by Earth with an ideal instrument.  The two NS masses in each binary of the catalogue are assigned by randomly sampling a uniform distribution between 1 and 2.5 $M_{\odot}$ \citep{abbottCBC}. Each NS has a dimensionless tidal deformability sampled from a uniform distribution between 0 and 2000. All BNS mergers in the catalogue are distributed isotropically in the sky with a random inclination of the orbital plane with respect to the line of sight. Given this catalogue of BNS mergers, \citeauthor{branchesi23} simulated one year of observations by ET operating at full sensitivity (xylophone configuration) in two possible geometries (triangular with 10 km arms and 2L with 15 km arms), assuming 85$\%$ uncorrelated duty cycle for each of the nested interferometers in ET. The analysis conducted with GWFish \citep{dupletsa23} produced the detector SNR and source parameters, along with relative errors for each source, including the sky-localisation uncertainty, $\Omega_{90}$, defined as the 90$\%$ credible region. Finally, they associated a KN light curve in the $ugrizy$ filters of the Vera Rubin Observatory (Rubin hereafter) with each merger detected with SNR$ > $8. All KN light curves were calibrated on AT2017gfo, using the model and the best-fit parameters from \cite{perego17} and properly including the effect on the light curves due to the inclination angle and cosmological and K-correction. For this reason, hereafter we will refer to this set of KN simulations as gfo-like KN sample.

Within this sample, we selected all the KN light curves corresponding to mergers detected with SNR$ > $8 and localised better than 100 $\deg^{2}$.
The arbitrary cut on sky localisation was chosen to limit the computation time due to the large number of events, and is based on the unlikeliness that optical telescopes will follow GW events with larger sky localisations.

We notice that this KN modelling, based on a black body approximation, leads to an overestimation of the UV component.
Therefore, when we build rough spectra from the photometry, as outlined in Section \ref{wstsim}, we correct the overestimation in the UV to correctly reproduce AT2017gfo. All the details of the correction are provided in Appendix \ref{appendix2}.

\subsection{BNS, GW detection and KN population simulations : Set 2 and 3} \label{theoretical_kn}

The original BNS merger population is reported in \cite{iorio23} and predicts a local merger rate $\mathcal{R}_{\mathrm{BNS}}$ = $107 \mathrm{Gpc^{-3}yr^{-1}}$ (common envelope parameter $\alpha = 1$). This merger rate is compatible with the $90\%$ credible interval $\mathcal{R}_{\mathrm{BNS}}$ $\in [10 - 1700] \mathrm{Gpc^{-3}yr^{-1}}$  reported in GWTC-3 \citep{abbottCBC}. Starting from this population, \citeauthor{loffredo25} built four catalogues of ten years of BNS mergers up to redshift z = 1, considering two possible NS mass distributions, i.e. a Gaussian and a uniform distribution, and two possible NS equations of state (EOSs), i.e. APR4 \citep{akmal98,baym71,douchin2001} and BLh \citep{bombaci18,logoteta21}. The Gaussian distribution is centred at 1.33$M_{\odot}$ with a standard deviation of 0.09 $M_{\odot}$ and reproduces the distribution of BNSs in our galaxy \citep{ozel12, kiziltan13,ozel16}. The uniform distribution ranges between 1.1 $M_{\odot}$ and a maximum mass $M_{\mathrm{max}}$ set by the NS EOS. Specifically, the maximum mass is 2.2 $M_{\odot}$ for the APR4 EOS and 2.1 $M_{\odot}$ for BLh. For each binary, the dimensionless tidal deformability was assigned by considering the NS mass and EOS.  As for Set 1, the mergers in the catalogues are distributed isotropically in the sky with a random inclination of the orbital plane. For each catalogue, \citeauthor{loffredo25} simulated ten years of observations by ET (xylophone configuration) in two possible geometries (triangular or 2L) operating alone (Set 2) or in a network with one CE interferometer (1CE, Set 3), providing the parameter estimation performed with GWFish \citep{dupletsa23} and the sky-localisation uncertainty obtained by each GW detector network. Then, they associated KN light curves in the $griz$ filter of Rubin with mergers detected with SNR$ > $8. The KN modelling assumes a two-component anisotropic KN ejecta. The ejecta properties are computed through numerical relativity-informed fitting formulas. Additionally, the model considers the effect on KN light curves induced by prompt collapse of the merger to black hole (BH), which is more likely for massive and asymmetric binaries. The model was calibrated to reproduce KN light curves compatible with AT2017gfo for BNS mergers with chirp mass and mass ratio similar to GW170817. However, it also takes into account diverse KN light curves associated with BNS possibly different from GW170817. KN light curves are informed with the properties of the NS EOS, predicting, in general, brighter KN light curves for the BLh EOS compared to the APR4 EOS. As for Set 1, the light curves are computed considering cosmological and K corrections. Finally, the catalogues from \cite{loffredo25} include the emission from the optical afterglow of short GRBs possibly produced during the mergers. The modelling of the jet is based on SGRB170817A \citep{ghirlanda19} and SGRBs observed so far \citep{ronchini22}. The optical afterglow is computed through the Python package \texttt{afterglowpy} \citep{ryan20}, using as parameters $\epsilon_{\mathrm{e}}$ = 0.1, p = 2.2, $\log_{10} \epsilon_{\mathrm{B}}$ uniformly distributed in the range $\bigl[ - 4, - 2 \bigr]$ and n $\in$ 0.25, 15 $\times$ 10$^{- 3}$ cm$^{- 3}$. This modelling reproduces the observed range of the optical fluxes of short GRBs \citep{kann11}. The fraction $f_{\mathrm{j}}$ of BNS mergers which can produce a GRB is still uncertain. Given a BNS merger population and a model of the GRB emission, $f_{\mathrm{j}}$ can be calibrated from the observed rate of GRBs \citep{ronchini22}. In our case,  $f_{\mathrm{j}}$ = 0.7 \citep{loffredo25, ronchini22} so that a GRB afterglow light curve is assigned only to 70$\%$ of the BNS mergers in our catalogue. 

From these catalogues, we select all the BNS mergers detected with SNR$ > $8 and localised better than 100 $\deg^{2}$ , for ET operating alone, and 40 $\deg^{2}$ , for ET operating in a network with 1CE, and the associated KN and SGRB afterglow light curves. 

The smaller threshold used for the network including ET and CE takes into account the larger number of relatively well-localised detections, and the limited observational time resources of optical telescopes.
To distinguish this dataset from the gfo-like KN one, we call hereafter the KN populations following these mergers the theoretical KN sample. 

Fig. \ref{mag_vs_z} shows the magnitude distribution of the optical counterparts presented in this section (Set 2 and Set 3) for a $2L$ configuration) at $\sim$12 hours post-merger.

\section{WST simulations} \label{wstsim}

\begin{figure*}[h!]
\centering
\includegraphics[width = 0.4\linewidth]{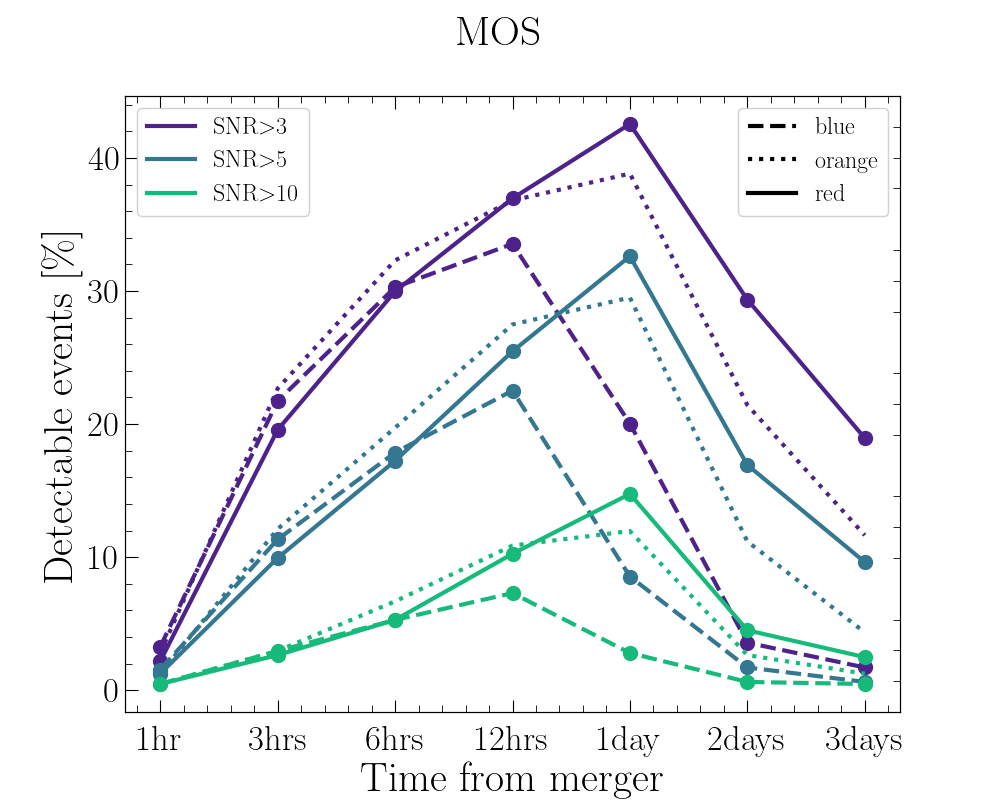}
\includegraphics[width = 0.4\linewidth]{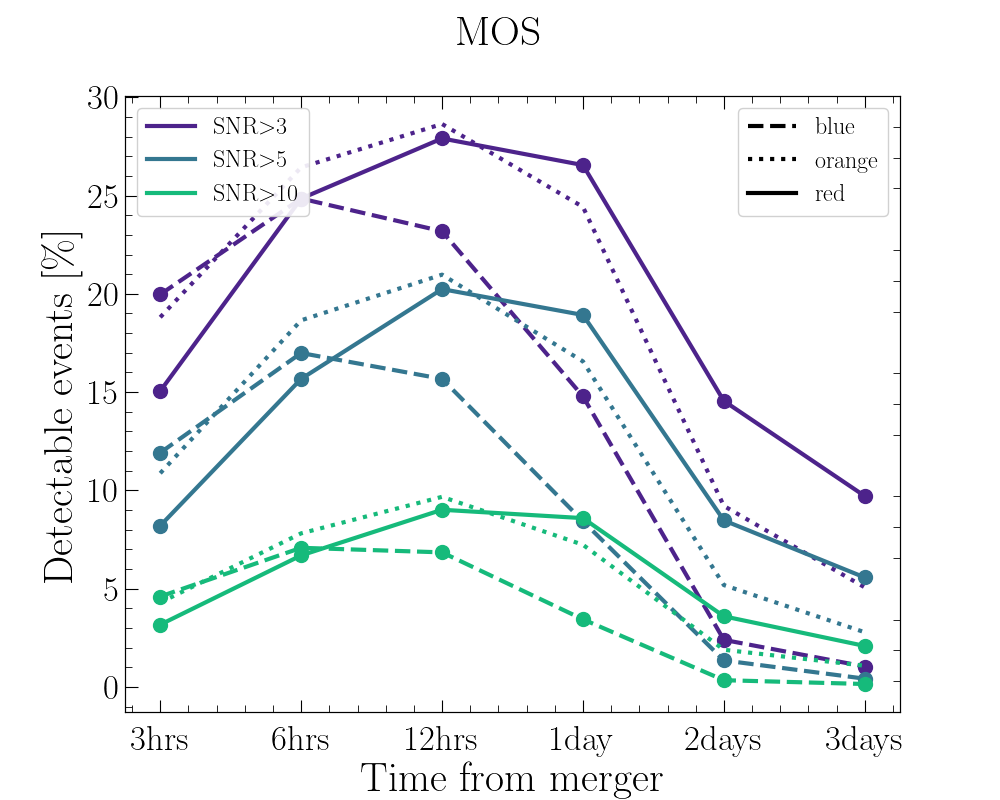}
\caption{Percentage of \textit{gfo-like} KNe (left) and \textit{theoretical} KNe (right) following BNS mergers detected by ET in the 2L configuration that are detectable with WST MOS. 
Different colors indicate different SNRs; dashed, dotted and solid lines correspond to the blue, orange and red arms of the spectrographs, respectively.
The highest detectability occurs approximately 12--24 hours post-merger.
}
\label{detectability_2L}
\end{figure*}

\begin{table}[ht]
\begin{center}
\small{
\begin{tabular}{|p{1cm}|p{2.5cm}|p{2.5cm}|p{1cm}|}
   
    \hline
    \multicolumn{4}{|c|}{WST} \\
    \hline
    channel & spectral range [$\AA$] & best throughput range [$\AA$]& spectel size [$\AA$] \\
    \hline
    & \multicolumn{3}{c|}{IFS} \\
    \hline
    blue & 3700-6100 & 4800-5800 & 0.64\\
    red & 6000-9600 & 6500-7500 & 0.97\\
    \hline
     & \multicolumn{3}{c|}{MOS} \\
    \hline
    blue & 3700-5350 & 4800-5300 & 0.41\\
    orange & 5150-7400 & 6000-7000 & 0.55\\
    red & 7200-9700 & 7300-8300 & 0.61\\
    \hline
\end{tabular}
}
\end{center}
\caption{Some characteristics of the WST IFS and MOS channels used in this study. The best throughput range is the one we chose to compute the SNR.}
\label{wst_info}
\end{table}

We perform simulations of WST observations to study the characteristics of the population that is {\it detectable} with WST, independently of any detailed operational strategy of WST observations in general, whose preparation and study are among the goals of this work. We provide the definition of {\it detectability} with WST, mainly based on a SNR threshold, later in this section.
For the purpose of this paper, we do not aim to simulate the KN spectral features, but the rough overall brightness of the continuum. This paper focuses on detectability of the EM counterparts with WST: a detailed analysis of spectral features is beyond the scope of this work.

For Set 2 and Set 3, \citeauthor{loffredo25} outline an observing strategy to adopt with Rubin to research the EM counterparts. According to this strategy, an event is classified as {\it detectable} with Rubin if it is observed in at least one epoch in both $g$ and $i$ filters, with exposures lasting 600 seconds. The observing strategy takes into account for the Observatory footprint and the sources position (RA, Dec) in the sky that are observed from the Chilean site of Cerro Pachón. To simulate WST detections within Set 2 and Set 3 we use the subsample of the KN and GRB afterglow populations that would be detectable by Rubin, considering that both WST and Rubin sites should be in the southern hemisphere, and the slightly deeper magnitude detectable by Rubin.

The simulations of WST observations were performed with the WST Exposure Time Calculator (ETC) \footnote{\url{https://github.com/RolandBacon/pyetc}}, properly modified to adapt it to this specific science case. In particular, we added to the Python source code functions to create artificial rough spectra, and to analyse different input spectra. With the ETC we simulated signal to noise ratio (SNR) per spectel (i.e. spectral pixel) for WST IFS and MOS low resolution spectrograph (see Table \ref{wst_info}) of KN and GRB afterglow emission following the BNS mergers in our samples.

We used simulated photometric data in Rubin filters of KN and GRB afterglows to build artificial rough spectra of each EM counterpart over the time. We started from the simulated AB magnitudes in Rubin filters - $g$, $r$, $i$ and $z$ for the {\it theoretical} sample, {\it ugrizy} for the {\it gfo-like} one - to which we added the contribution of Galactic dust extinction. Then we converted them into observed flux density ($\mathrm{erg\,s^{-1} \, cm^{-2} \, \AA^{-1}}$) at each filter effective wavelength, approximately considered as the medium flux in that filter. Thus, we obtained a dataset of 4 (6) flux density values at 4 (6) different wavelengths for each event at each time step: the fit of these values with a quartic (sextic) polynomial was used as input spectrum of WST simulations.
Once we had rough spectra both for the {\it gfo-like KN} and {\it theoretical KN} samples, we computed SNR over the best throughput wavelength range of WST IFS and MOS channels. The SNR of MOS was spatially integrated over the fibre aperture (1 $\arcsec$ assumed in version 0.31 of the ETC), while for the IFS we used a specific function of the ETC, computing the optimum aperture diameter that maximises the SNR.

We consider an event {\it detectable} when the average SNR in the selected spectral range exceeds the threshold of 3. We assume a seeing of $0.7\arcsec$ and an airmass of 1 as observing conditions. These are optimal observing conditions, but a detailed analysis goes beyond the scope of this work.
In the sections that follow, we show and discuss the properties of the BNS populations and their detectable counterparts, sampled at approximately 3, 6, 12, 24, 48 and 72 hours (observer frame) after the merger. We consider exposures lasting 1800\,s and 3600\,s: we present in the main section results for 3600\,s exposures, and in Appendix \ref{appendix} for 1800\,s.

\section{Results}\label{results}

The results of the work presented in the following sections do not depend on the configuration chosen for ET. Therefore, we decided to show only those based on the simulation of the population detected by the $2L$ design. 

Similarly, we do not observe significant differences in redshift horizon, median magnitudes, detectability trends over time, or other observable properties relevant to assessing the impact of IFS and MOS with WST, across the four combinations of NS EoS and mass distributions considered. Differences do emerge in the total number of detections, but they align with those already discussed in \cite{loffredo25}, without introducing additional dependencies. Therefore we show results for the BLh EoS and the gaussian mass distribution only.

\subsection{Detectable fraction and best observing time post trigger}

The percentage of detectable EM counterparts primarily depends on time elapsed since the merger and the instrument used (with IFS being slightly more performative in this regard with respect to fibres).
Figure \ref{detectability_2L} shows the detectable percentages of {\it gfo-like} and {\it theoretical KN} as a function of time, using WST fibres. Results for the IFS are shown in Appendix \ref{appendix3} (Fig. \ref{detectability_2L_IFS}). 
As expected, KNe become fainter in the $g$ band with respect to the $i$ band at late epochs (see also Fig.\,\ref{mag_i_vs_mag_g}), hence the detectability improves in the redder bands. 
The best time to trigger observations appears to be from about 12 hours to 1 day after the merger, as this is when the highest detectability is achieved.

\begin{figure}
\centering
\includegraphics[width = 8.3cm]{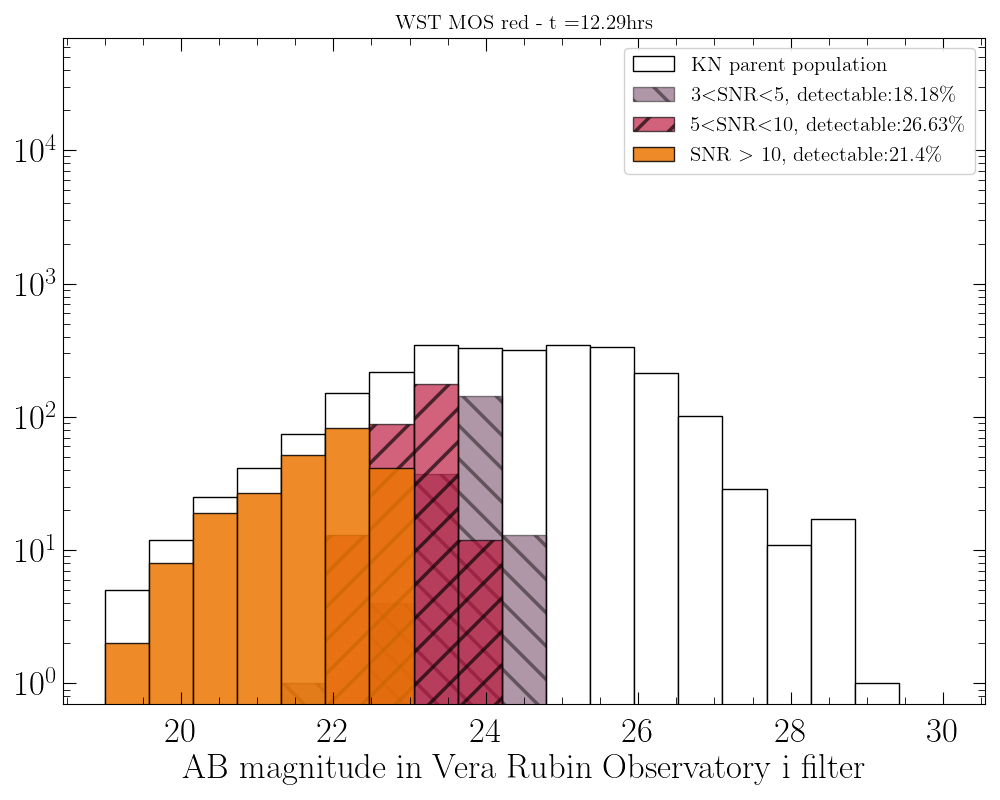}
\caption{Magnitude distribution in $i$ filter of ET BNS detected over 10 years and their corresponding \textit{theoretical KN} at $\sim$ 12 hours after the merger.  
The background distribution in white corresponds to the parent BNS+KN population. 
The colored distributions correspond to the EM counterparts that are detectable with WST MOS red arm, different colors corresponding to different SNR intervals. }
\label{mag_i_distribution}
\end{figure}

In Fig. \ref{mag_i_distribution} we show the fraction of the WST MOS detectable events at 12 hours after the BNS merger as a function of their magnitude. Results for the IFS are reported in Appendix \ref{appendix3} (Fig. \ref{mag_i_distribution_IFS}). With WST it will be possible to detect KNe having $i$-band magnitudes as faint as $\sim$ 25 with fibres and $\sim25.5$ with the IFS (see Appendix \ref{appendix3}, Fig. \ref{mag_g_distribution}, for the analogous Figure for the $g$ band). The bright sources that are not detected, are mainly missed because not
in the sky-coverage of VRO and WST.

\subsection{Properties of the KN population detectable at 12 hours post-merger}

\begin{figure}[h!]
\centering
\includegraphics[width = 9cm]{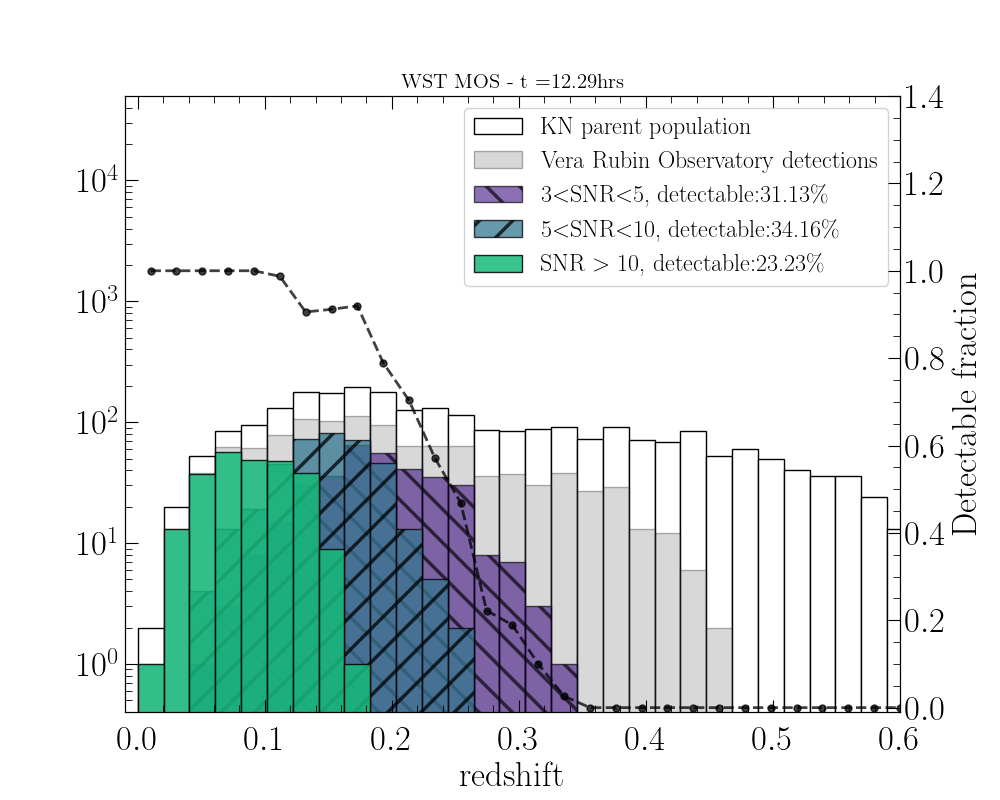}
\caption{Redshift distribution of ET BNS detected over 10 years and the corresponding \textit{theoretical KN} at $\sim$ 12 hours after the merger.
The background distribution in white corresponds to the parent BNS+KN population. 
The colored distributions correspond to the EM counterparts that are detectable with WST MOS blue or orange or red arm, different colors corresponding to different SNR intervals. EM counterparts detectable with Rubin are shown in grey. Black points refer to the y axis ticks on the right hand side and show the fraction of detectable (SNR $>$ 3) EM counterparts, with no additional information on the SNR range, with respect to Rubin detections.}
\label{z_distribution}
\end{figure}

Figure~\ref{z_distribution} shows the redshift distribution of ET BNS events and their corresponding \textit{theoretical} KNe approximately 12 hours post-merger, along with the distribution of KNe detectable using WST fibres. The detection limit with fibres is around $z \sim 0.35$, while the IFS allows detection up to $z \sim 0.4$ (see Appendix~\ref{appendix3}, Fig.~\ref{z_distribution_IFS}, for the redshift distribution using WST IFS). 

A crucial factor that will influence the choice of the observing strategy, and the real chances for these electromagnetic counterparts to be observed, is the sky-localisation uncertainty, $\Omega_{90}$.

\begin{figure*}
\centering
\includegraphics[width = 8cm]{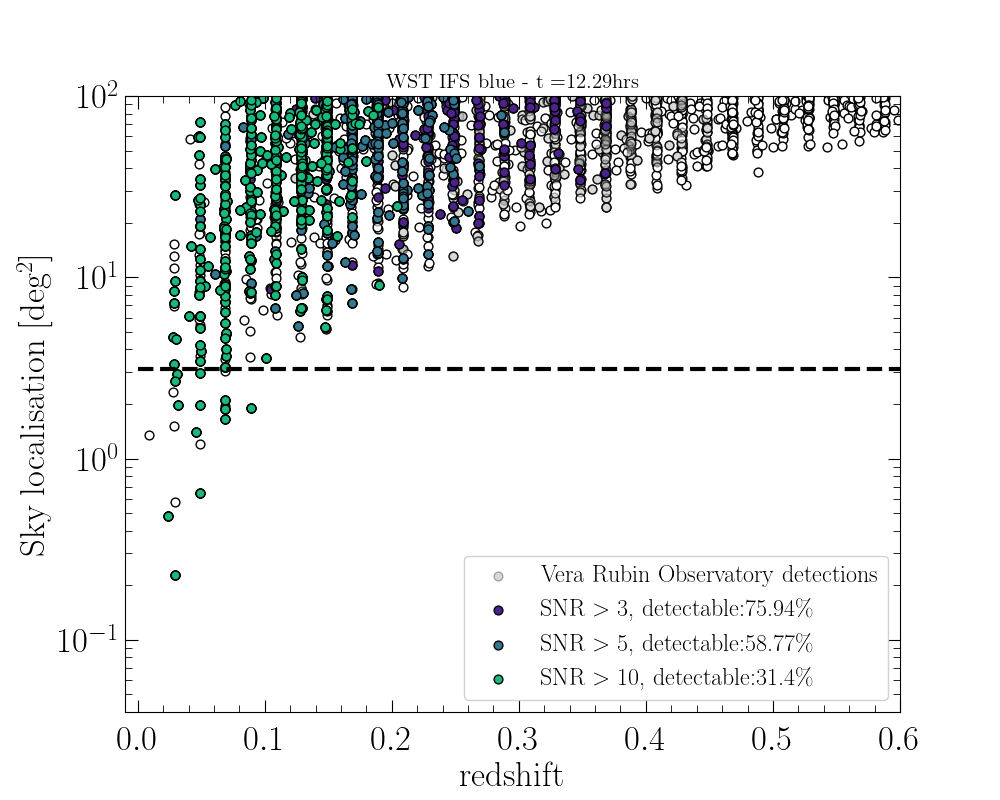}
\includegraphics[width = 8cm]{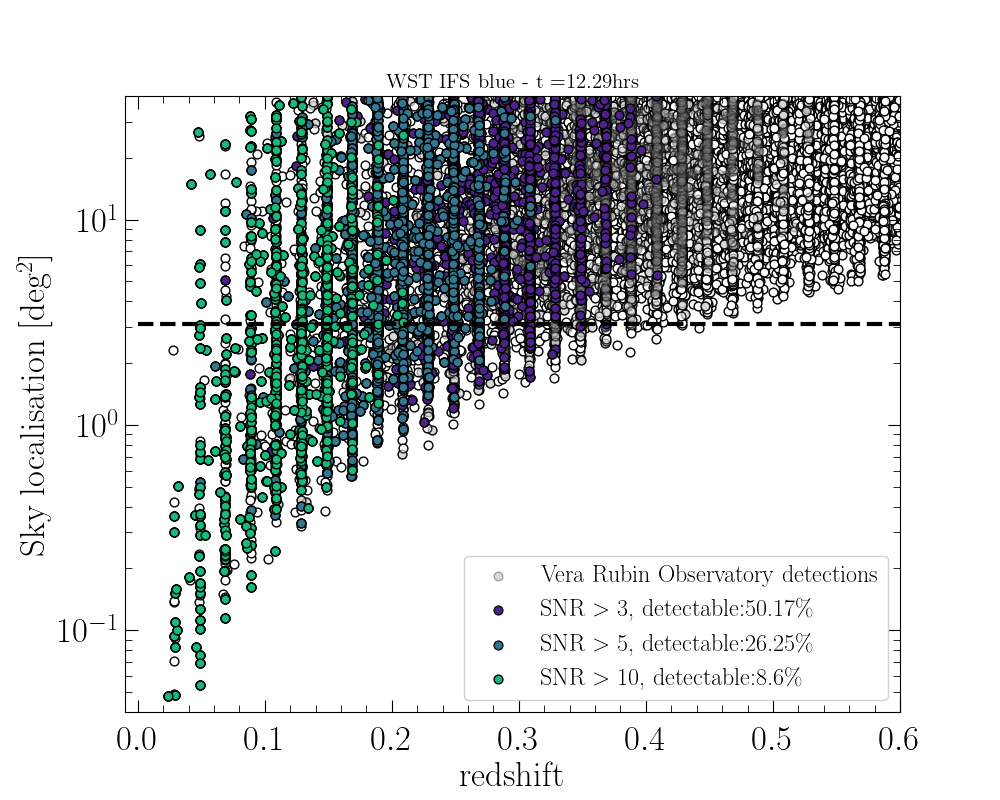}
\caption{Sky localisation uncertainty as a function of redshift of BNS (white points) detected with ET (left panel) and ET+CE (right panel) over 10 years. Grey points refer to the EM counterparts detectable with Rubin, among which the colored points refer to those that are detectable with WST IFS blue arm at $\sim$ 12 hours post-merger. Different colors correspond to different SNR. Percentage reported in the legend refer to the fraction of EM counterparts that are detectable with WST, with respect to those that are detectable with Rubin. The black dashed line represents WST patrol area, 3.1 deg$^{2}$, where fibres will be positioned.}
\label{ETvsETCE_skyloc}
\end{figure*}

$\Omega_{90}$ of BNS detected by ET alone rapidly worsens with increasing redshift and covers regions as small as a fraction of $\deg^{2}$, up to the maximum limit of $100 \deg^{2}$ that we imposed. A network including ET and CE benefits from improved sky localisation at higher redshifts and results in a significantly higher number of well localised events at lower redshift (Fig.\ref{ETvsETCE_skyloc}).

\begin{figure}
\centering
\includegraphics[width = 8.8cm]{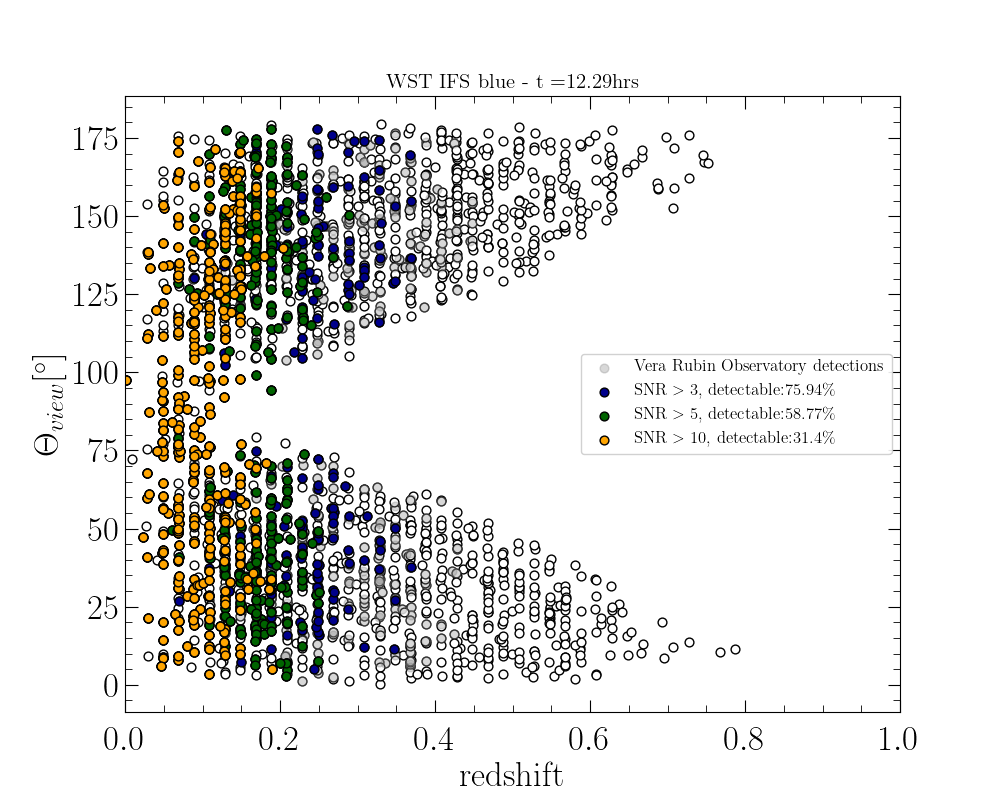}
\includegraphics[width = 8.8cm]{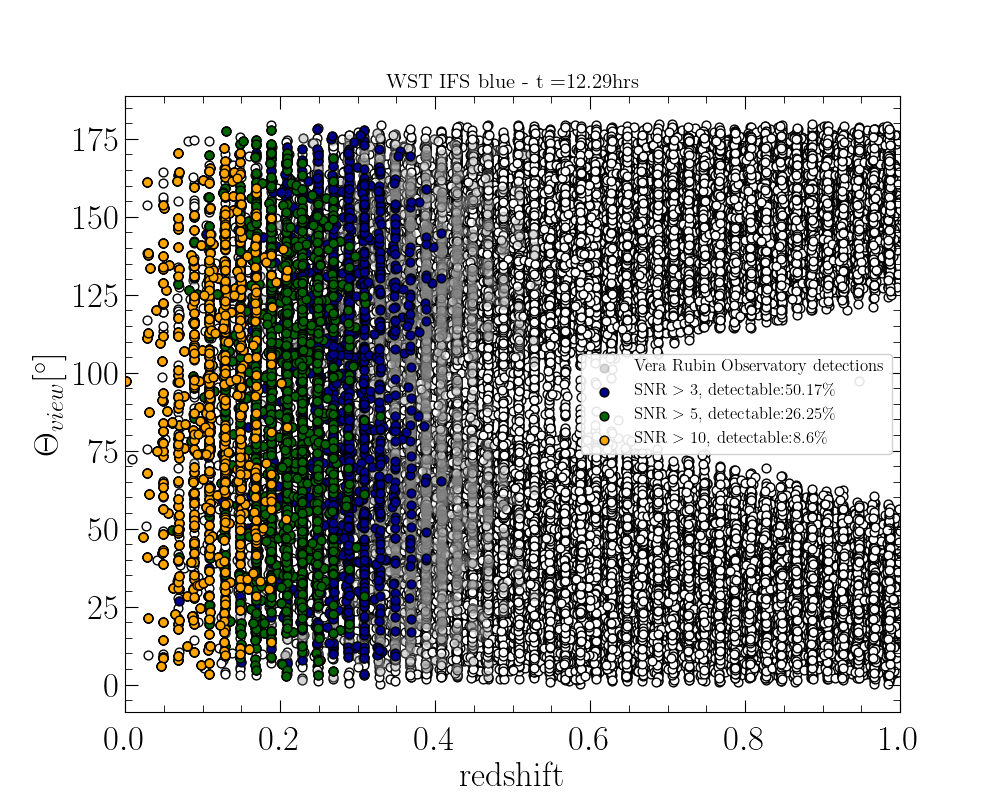}
\caption{Viewing angle as a function of redshift of \textit{theoretical KN} following BNS mergers (white points) detected by ET alone (top panel) and ET in a network with CE (bottom panel).
Colored points represent WST detections, different colors corresponding to different SNR. The lack in GW detections for edge-on systems is due to GW signal SNR, that is higher for face-on systems with respect to edge-on systems with the same intrinsic parameters.}
\label{theta_vs_z}
\end{figure}

The last property we explore here is the viewing angle $\Theta_{\mathrm{view}}$ of the BNS, that is the angle between the BNS angular momentum axis and the line of sight. Figure \ref{theta_vs_z} shows the population of BNS detected by ET alone and in a network with CE, highlighting those whose counterpart is detectable with WST. As redshift increases, there is a noticeable absence of edge-on systems. This happens because the SNR of the GW signal is higher for face-on systems compared to edge-on systems with the same intrinsic parameters. 

 When considering ET alone, WST does not detect edge-on systems beyond $z\sim0.1$. However, this limitation is only due to the redshift distribution of the GW-detected BNS population with respect to the viewing angle. 

The dependence of sky localisation on the viewing angle is relatively weak for BNS GW detections, although it tends to worsen for edge-on systems due to their lower SNR. Therefore, large FoV and high multiplexing facilities can be necessary for the spectroscopic follow-up independently of the viewing angle of the BNS event (see, however, Sect. \ref{grbafterglows} for the case of well localised on-axis GRB afterglows). 

\subsection{GRB afterglows contribution}\label{grbafterglows}

\begin{table}
    \centering
    \begin{tabular}{|p{3cm}|p{1cm}|p{1cm}|p{1cm}|}
        \hline
        & 3 h & 12 h & 3 d  \\
        \hline
        $\Theta_{view} < 10 ^{\circ}$  & 99 $\%$ & 93 $\%$ & 95 $\%$ \\
        \hline
        $10 ^{\circ} < \Theta_{view} < 15 ^{\circ}$ & 97 $\%$ & 62 $\%$ & 78 $\%$ \\
        \hline
        $15^{\circ} < \Theta_{view} < 20 ^{\circ}$ & 30 $\%$ & 8 $\%$ & 17 $\%$ \\
        \hline
    \end{tabular}
    \caption{Percentage of afterglows outshining KNe in different bins of viewing angle $\Theta_{\mathrm{view}}$, at different times post-merger. The percentages are given with respect to the population of events detectable with WST IFS red arm and that feature an afterglow.}
    \label{KN_AG}
\end{table}

\begin{figure}
    \centering
    \includegraphics[width=0.85\linewidth]{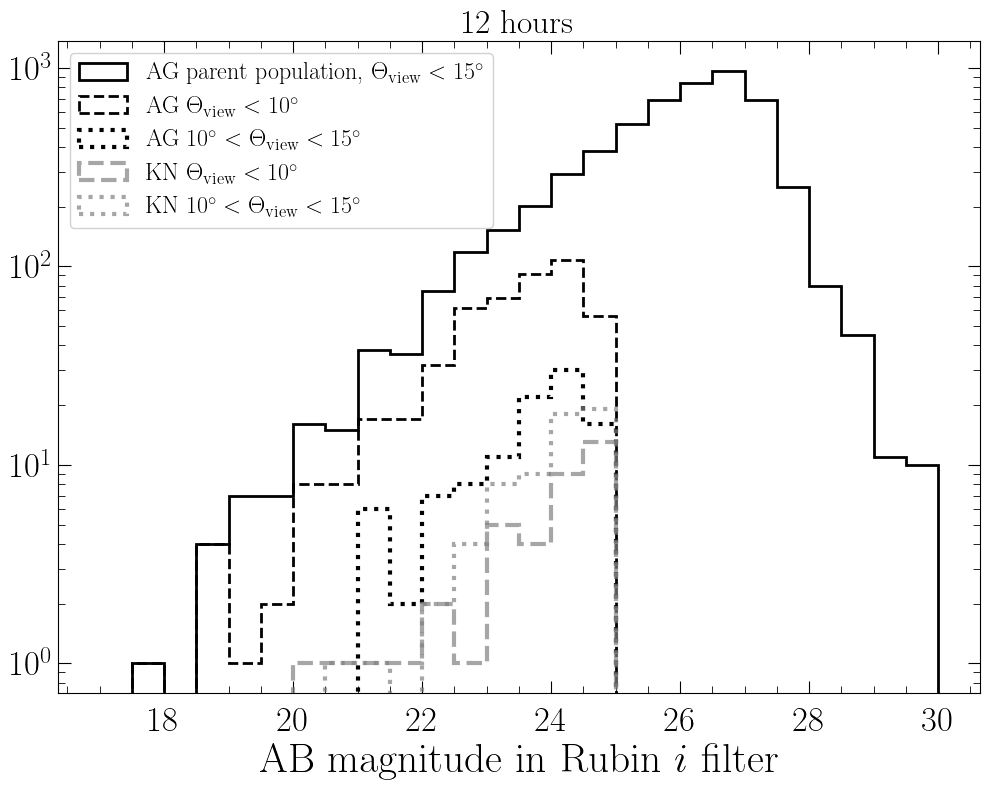}
    \includegraphics[width=0.85\linewidth]{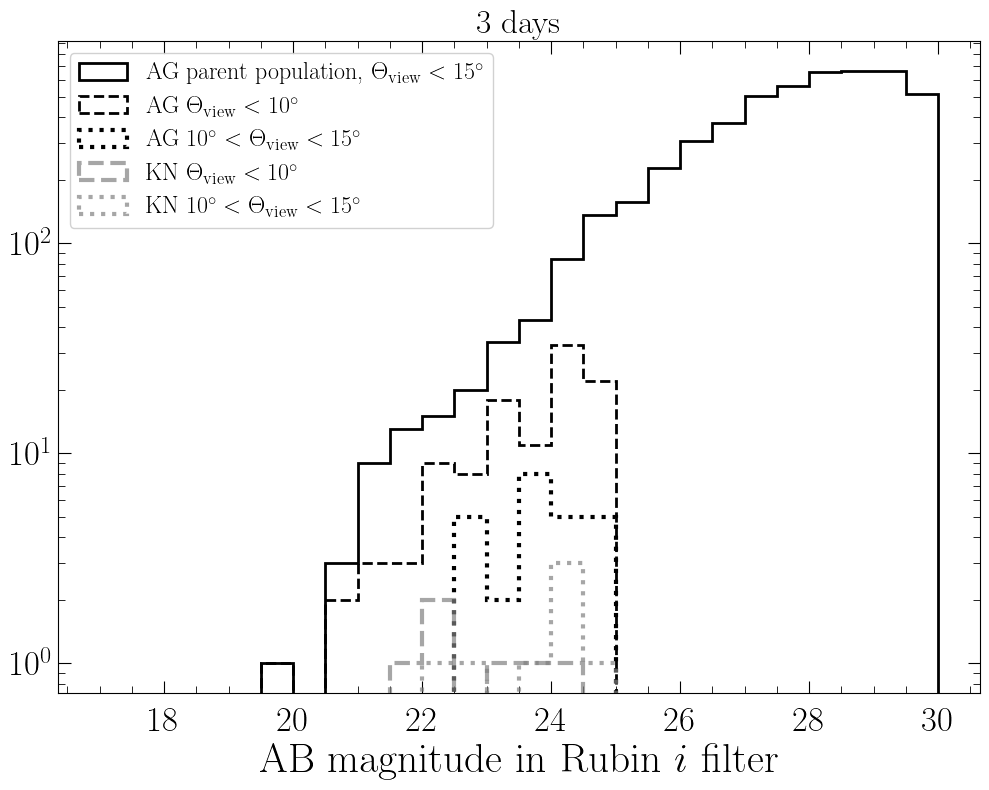}
    \caption{Magnitude distribution in the Rubin $i$ filter (with a cut at $m = 30$) of afterglows produced by BNS mergers with a viewing angle below 15$^{\circ}$ (solid black line), at $\sim$12 hours (top panel) and $\sim$3 days (bottom panel) post-merger. Dashed- and dot-line black histograms correspond to events that are detectable with WST (i.e., where the combined flux of the afterglow and KN has $\mathrm{SNR} > 3$) and for which the afterglow outshines the KN, for $\Theta_{\mathrm{view}}<10^{\circ}$ and $10^{\circ} < \Theta_{\mathrm{view}}<15^{\circ}$, respectively. Dashed- and dot-line gray histograms represent detectable events  where the kilonova dominates over the afterglow, for $\Theta_{\mathrm{view}}<10^{\circ}$ and $10^{\circ} < \Theta_{\mathrm{view}}<15^{\circ}$, respectively.}
    \label{mag_ag}
\end{figure}

GRB afterglows (on- or off-axis) are one of the expected EM counterparts from BNS mergers. We study the afterglow detectability by WST.
We used the afterglow simulations reported in Sect. \ref{theoretical_kn}, and considered different observing times from 3 hours to 3 days post-merger, for BNS detected by ET in a network with CE. We find that the contribution of the afterglows is mainly limited to on-axis and slightly off-axis systems ($\Theta_{\mathrm{view}} < 15^{\circ}$; see Table \ref{KN_AG} and Fig. \ref{theta_agkn}). 
More than $90\%$ of the events with $\Theta_{\mathrm{view}} < 10^{\circ}$ have afterglows that outshine the KN and are detectable with WST (see Fig.\,\ref{mag_ag} for the 12 hours and 3 days cases), at any of the considered times. The fraction decreases significantly for larger viewing angles and later observations, becoming null at $\Theta_{\mathrm{view}}\sim20^{\circ}$. 

We stress that, contrary to KN, afterglows with $\Theta_{\mathrm{view}}\sim15^{\circ}$ can be detected by WST (see also Fig. \ref{theta_agkn}) also for events beyond $z=0.4$ (even beyond $z=1$). For such cases, the possibility of a detection is optimized by observations starting as soon as possible after the merger, because the optical lightcurve of on-axis afterglows peaks at earlier times than KN ones. Our simulations show that at 12 hours the fraction of detectable afterglows with $\Theta_{\mathrm{view}}\sim15^{\circ}$ is reduced by half compared to the case of observations starting 3 hours after the merger.

Spectroscopic follow-up with WST is well suited for both on-axis and off-axis systems, albeit with some differences. The use of WST large number of fibres to identify the EM counterpart of an on-axis BNS will not be necessary when the GRB prompt emission is already detected by satellites capable of detecting and providing rapid precise localisations of GRBs and their afterglows (supposing that such satellites will be operational in the 2040s). However, deploying the IFS on the GRB localisation remains extremely valuable in case of localisations spanning several arcminutes to obtaining spectra of the field, besides those of the counterpart itself. This enables the measurement of redshifts for galaxies in the field, particularly relevant for BNS systems with significant offsets from the centres of their host galaxy (see Sect. \ref{challenges_galaxy_targeted}), and facilitates both the identification of the host galaxy and the spectroscopic characterisation of the counterpart itself.
For the case of slightly off-axis GRBs and afterglows ($\theta_{\mathrm{jet}}<15^{\circ}$), where a precise enough localisation may not be available, WST high multiplexing capability becomes crucial for identifying the EM counterpart.

\section{Observing strategy}\label{obs_strategy}
We consider two possible observing strategies to adopt with WST to search EM counterparts of BNS GW signals from next generation observatories. We outline them in the following section, where we also investigate their feasibility, the possible issues and the expected outcomes. For both strategies, telescope-level ToOs are considered.

The first strategy consists in exploring the GW signal error region with WST alone performing a galaxy targeted research, possibly covering the peak of the probability regions with the IFS and the remaining area with fibres. The second strategy that we consider is to use WST in synergy with optical-NIR photometric observations, targeting the EM counterpart candidates that they will provide.

In this work we only consider galaxy selection made on redshift and not based on additional properties such as stellar mass and star formation rate (SFR), as this is beyond the scope of our study. We note that, despite selections based on high values of stellar mass or SFR have been used for the searches of EM counterparts of BNS detected by LVK, those are not the typical properties of SGRB host galaxies, that span different values of star formation rate and stellar mass  \citep{nugent22}.

\subsection{Galaxy-targeted research in a stand-alone scenario}\label{n_galaxies}

In the perspective of performing galaxy-targeted searches of the EM counterparts within the 3D error volume of the GW signal, we aim to give an estimate of the number of galaxies that can be found within such volumes.
For this exercise, we suppose that the relevant information on the photometric or spectroscopic redshift of galaxies will be available from previous surveys or from the WST galaxy surveys.

First, we select the BNS with $z<0.45$, as this corresponds to the detection horizon of the potentially associated  KN. We do not apply here any further pre-selection based on which events are detectable with Rubin or WST.
Then we estimate the 3D error region of each of them as follows. We consider the differential comoving volume:

\begin{equation}
\centering
    \hspace{1cm}\frac{d^{2}V_{\mathrm{C}}}{d\Omega dz} = D_{\mathrm{H}}\frac{(1+z)^{2}D_{\mathrm{A}}^{2}}{E(z)} 
\end{equation}
\label{dVc}

where $E(z)$ and $D_{\mathrm{H}}$ are defined as

\begin{gather*}
    E(z) = \sqrt{\Omega_{\mathrm{M}}(1+z)^{3} + \Omega_{\mathrm{K}}(1+z)^{2} + \Omega_{\mathrm{\Lambda}}}\\
    D_{\mathrm{H}} = \frac{c}{\mathrm{H_{0}}}
\end{gather*}

We multiply the differential comoving volume by the uncertainty on the sky localisation of the BNS (2D error area) and we integrate this product over redshift to obtain the 3D error region, that we call hereafter {\it comoving error volume} ($V$). To define the bounds of the integral we convert $D_{\mathrm{L}}$ - $\Delta D_{\mathrm{L}}$ and $D_{\mathrm{L}}$ + $\Delta D_{\mathrm{L}}$ into redshift values ($z_{1}$ and $z_{2}$), where $D_{\mathrm{L}}$ is the luminosity distance estimated from the GW signal and $\Delta D_{\mathrm{L}}$ is its associated error.
If $\Delta D_{L} > D_{L}$ we set z$_{1}$ = 0 as lower bound.
 
We note that on average, sky localisation and the error on $D_{\mathrm{L}}$ worsen with redshift, but we stress that large error volumes containing a huge number of galaxies can be found even at low redshift (See Fig. \ref{comoving_volume_skyloc} and \ref{skylocDL}).

\begin{figure}[h!]
    \includegraphics[width = 0.5\textwidth]{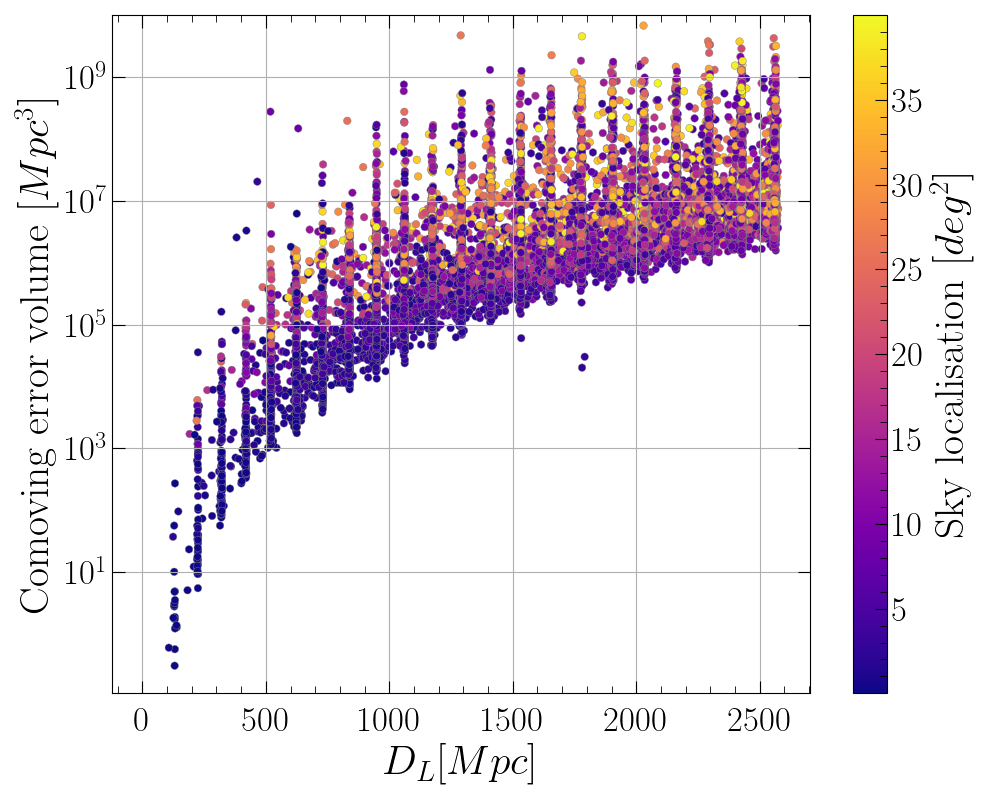}
    \caption{Comoving error volume as a function of GW luminosity distance for each BNS detected by ET in a network with CE at $z<0.45$. The color of the points refers to the uncertainty on the sky localisation, according to the color bar on the right.}
    \label{comoving_volume_z}
\end{figure}

\begin{figure}
    \centering
    \includegraphics[width = 0.4\textwidth]{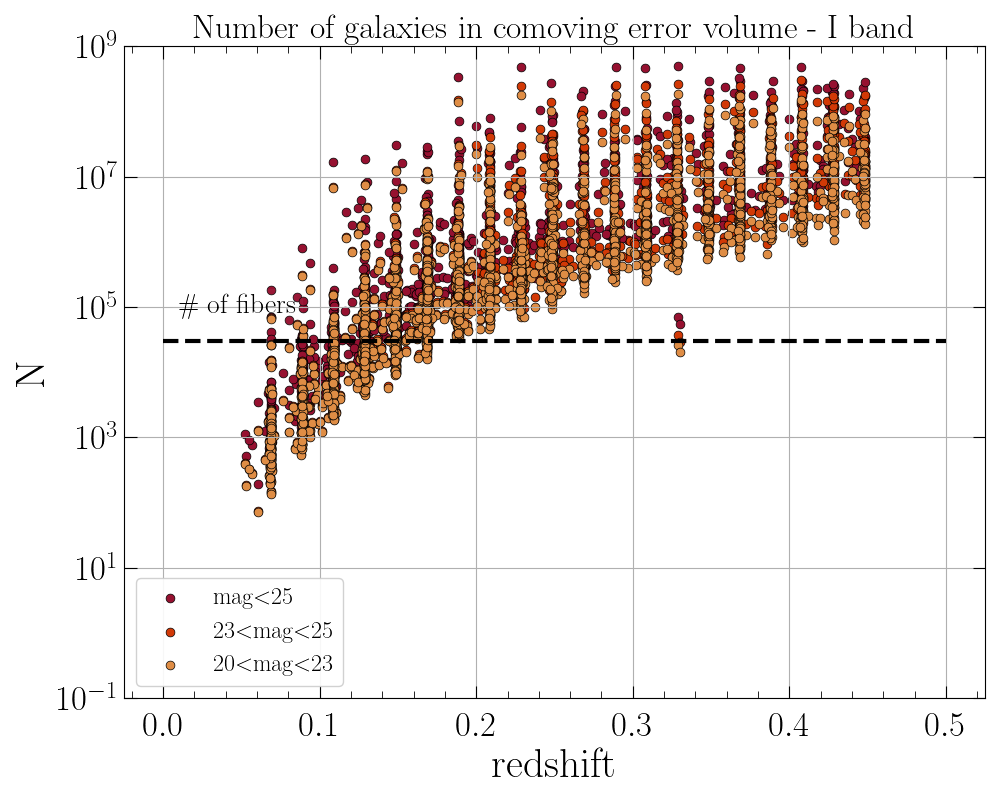}
    \includegraphics[width = 0.4\textwidth]{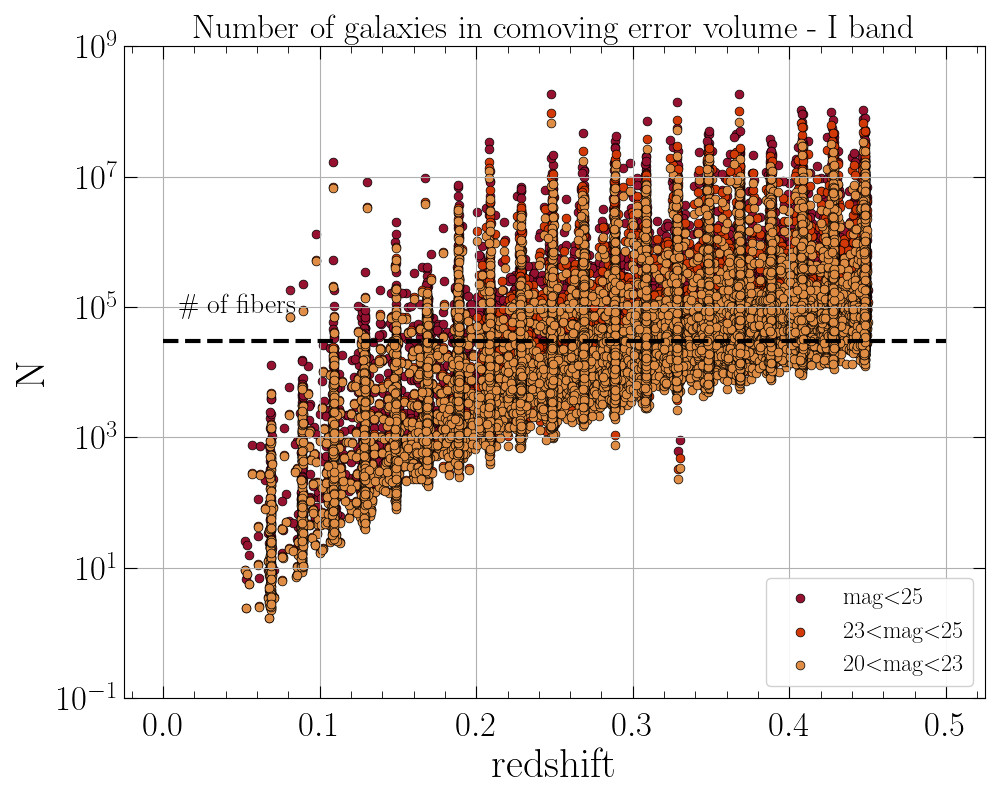}
    \includegraphics[width = 0.4\textwidth]{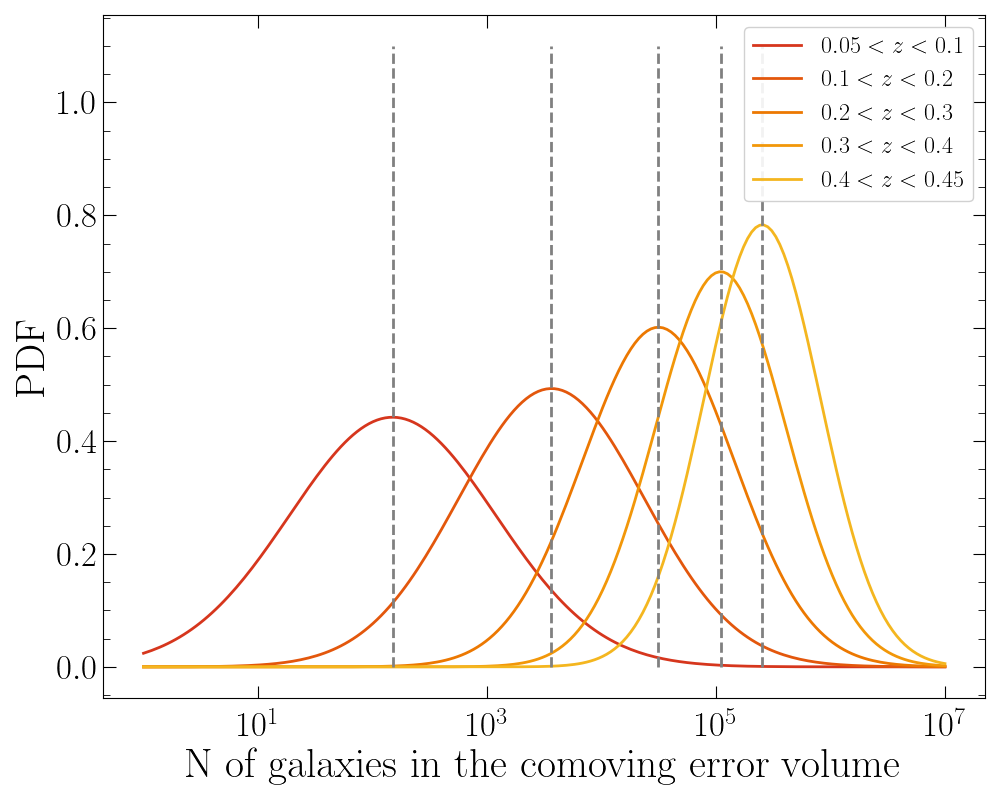}
    \caption{Top: Number of galaxies in the comoving error volume of BNS detected by ET alone (top panel) and in a network with CE (middle panel). Each point corresponds to one GW BNS detection. Different colors correspond to different magnitude intervals. The black dashed line represents the number of WST fibres.
    Bottom: Probability distribution function of the number of galaxies with apparent $I$-band magnitude $m < 25$ that can be found in the comoving error volume of each BNS detected by ET+CE, for different redshift intervals. Dashed vertical lines indicate the mean value around which each Gaussian is centred.}
    \label{ilbert_I}
\end{figure}

To estimate the number of galaxies within the comoving error volume of each BNS we multiply such volume by galaxy number density, computed by integrating the galaxy luminosity function (GLF) in different magnitude intervals.

The basic shape of GLF is given, in terms of absolute magnitude $M$, as a single Schechter function:

\begin{equation}
    \Phi(M) = 0.4 \cdot ln(10) \cdot \Phi^{*} \cdot 10^{0.4(\alpha + 1)(M^{*} - M)} exp \left[-10^{0.4(M^{*}-M)}\right]
\end{equation}

where $M^{*}$ is the characteristic absolute magnitude, $\alpha$ represents the faint-end slope parameter and $\Phi^{*}$ is a normalisation constant giving the number of galaxies per Mpc$^{3}$ \citep{schechter76}.
$\Phi(M)$ can be written in terms of the apparent magnitude m as follows:

\begin{equation}
\begin{split}
    \Phi(m) = 0.4 \cdot ln(10) \cdot \Phi^{*} \cdot 10^{0.4(\alpha + 1)(M^{*} - m + 5log_{10}(D_{\mathrm{L}}) - 5)} \times  \\
    \times exp \left[-10^{0.4(M^{*} - m + 5log_{10}(D_{\mathrm{L}}) - 5)}\right]
\end{split}
\end{equation}

where D$_{\mathrm{L}}$ is the luminosity distance in parsec.
We assume that once D$_{\mathrm{L}}$ is fixed, $\Phi(m)$ gives the number of galaxies per unit apparent magnitude per unit volume [Mpc$^{-3}$] at that D$_{\mathrm{L}}$, and remains nearly constant in redshift across the comoving error volume. Hence for each event we integrate $\Phi(m)$ over different magnitude intervals. Then we multiply the number density of galaxies in each magnitude bin by the comoving error volume.
We use the Schechter function parametrisation from \cite{ilbert05} for redshift $0.05 < z < 0.5$.

\begin{figure}
    \centering
    \includegraphics[width = 0.5\textwidth]{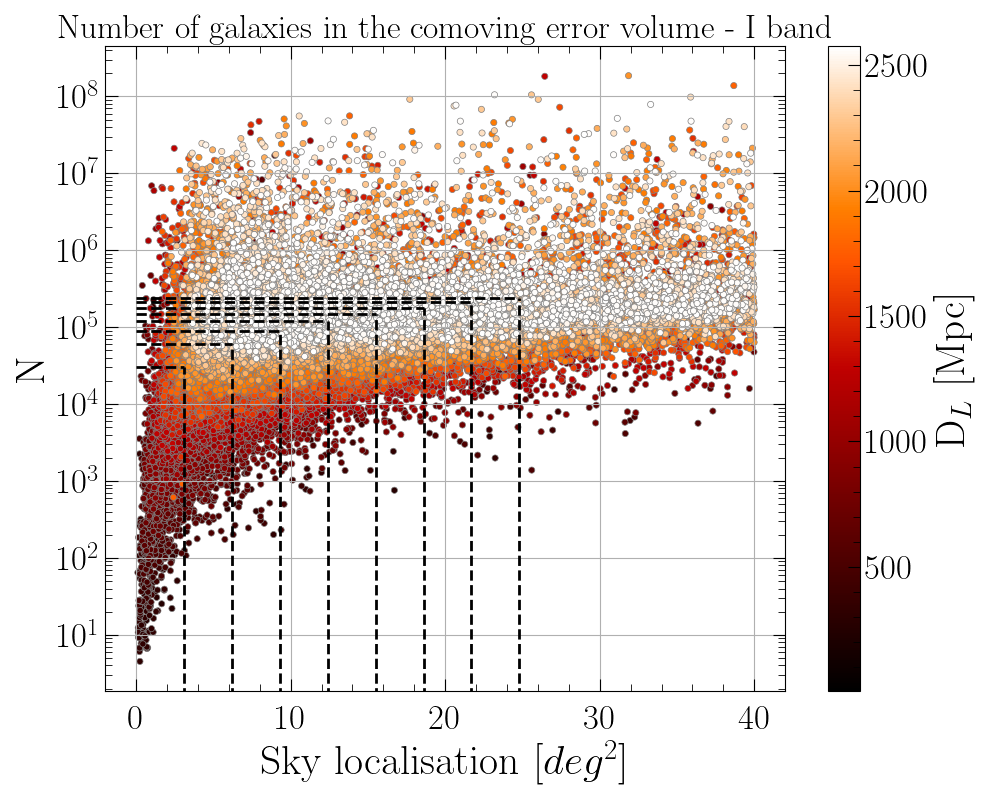}
    \caption{Number of galaxies in the comoving error volume of BNS detected by ET in a network with CE, as a function of their sky localisation. Each point represents one GW BNS detection. The black dashed lines represent an indication of what WST can cover in one night of observations, as described in section \ref{n_galaxies}.
    }
    \label{ilbert_I_skyloc_60}
\end{figure}

Figure \ref{ilbert_I} shows the number of galaxies within the comoving error volume of the BNS detected by ET observing alone and in a network with CE as a function of redshift. The number density is derived from the GLF in the I band. Fig. \ref{ilbert_I_skyloc_60} shows the number of galaxies within the comoving error volume of the same events, as a function of sky localisation. 
Also, if we initially assume that each galaxy will be targeted with $\sim$ 1 fibre, we can compare the number of galaxies with the number of WST fibres that can potentially be used. 
If the 2D error region of the GW event is tiled without overlaps, the portion of the region that can be covered is given by $n_{\mathrm{exposure}} \cdot \mathrm{FoV_{WST}}$, which is $\sim n_{\mathrm{exposure}} \cdot 3.1 \deg^{2}$. We assume that one night lasts $\sim$ 8 hours, so 8 exposures of 1 hour each can be performed. The dashed vertical lines on the plot indicate the size of the sky region that can be covered for an increasing total number of exposures. For each of these vertical lines, a corresponding horizontal line indicates on the y-axis the number of fibres that can potentially be used to target the galaxies within the comoving error volume with WST, that corresponds to $n_{\mathrm{exposure}} \cdot 30\,000$, considering the total number of fibres that are planned for this facility.
We note that a subset of events at intermediate distances corresponds to cases with $N > 10^6$ galaxies within the comoving error volume. These are primarily systems with viewing angles $\Theta_{\mathrm{view}} < 30^{\circ}$, for which the distance–inclination degeneracy is particularly difficult to break using GW data alone, resulting in large uncertainties on $\Delta D_{\mathrm{L}}$ (see Fig. \ref{ilbert_I_skyloc_ETCE_coltheta}). The same Figure for BNS detected by ET as a standalone observatory is reported in Appendix \ref{appendix3} (Fig. \ref{ilbert_I_skyloc_ETonly}).

The right balance must be found in the choice of the exposure time to adopt with WST, between long exposures allowing to perform deeper observations, and short exposures allowing to tile larger regions of the sky. Overall, in a stand-alone scenario where WST is used to target galaxies without additional information on EM counterpart candidates from previous photometric observations, shorter ($\sim$30-minute exposures) are preferable.
In this framework, the high sensitivity of the IFS can be leveraged to primarily target the peak of the probability distribution within the GW sky localisation and the regions containing extended galaxies.

\subsection{Further investigation of size and magnitude of galaxies in the comoving error volume} \label{challenges_galaxy_targeted}

A galaxy-targeted search is a way to find the EM counterpart from a system gravitationally bounded to its host galaxy. However, there are some feasibility caveats that have to be taken into account, especially for fibre observations, concerning the host galaxy sizes, the counterpart brightness and the counterpart offset from the host galaxy centre.

Except for on-axis GRB afterglow detections, which can outshine the host galaxy, the EM counterparts of BNS mergers are typically faint and may be outshined by their host galaxies. 
Typical magnitudes of galaxies at $z<0.5$ \citep{ilbert05}, as well as those of the host galaxy of SGRB in the same redshift range reported in the {\it Broad-band Repository for Investigating Gamma-ray burst Host galaxies Traits} (BRIGHT, \cite{fong22,nugent22}), are brighter than the distribution of the median magnitude of KN and GRB afterglows following ET BNS. 
However, these counterparts are not expected to be found at small offsets from their host centre, as also testified by the measured SGRB projected physical offsets. The median value of SGRB offsets is of 1.5\,$r_e$ \citep{fong22} (where $r_e$ is the effective radius of the host galaxy), with more than 60\% of those at $z<0.5$ lying at $>1 r_e$. Considering the host galaxy profiles of the SGRB at $z<0.5$ presented in the \cite{Fong13} sample, the host galaxy surface brightness at the offsets of the SGRB would be comparable or much fainter than those expected for KN detectable by WST. In such cases, the possibility of identifying the KN spectrum overimposed on the host galaxy one would not be compromised.

\begin{figure}
    \centering
    \includegraphics[width=0.9\linewidth]{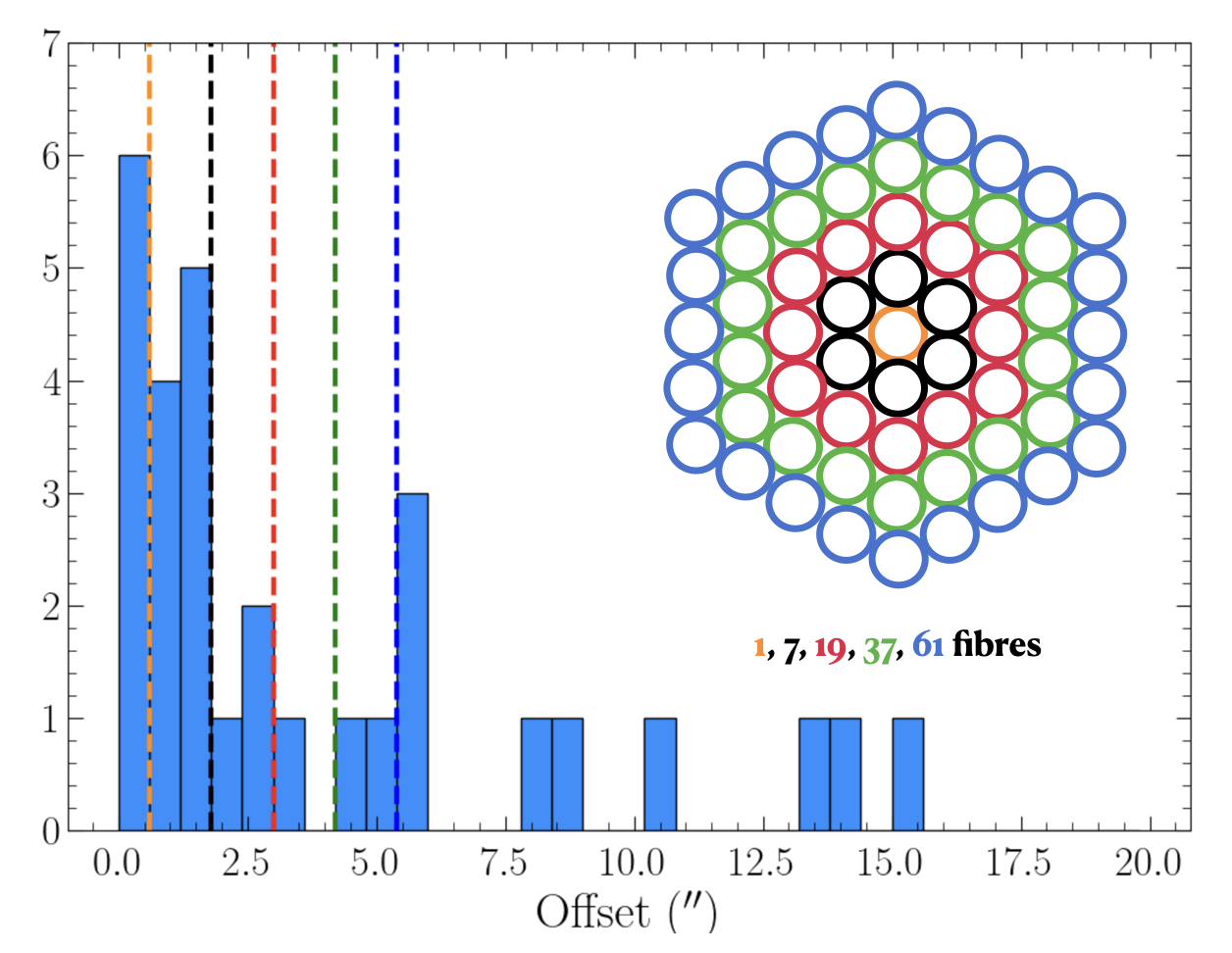}
    \caption{Distribution of the offsets of SGRBs at $z < 0.5$ from the sample presented in \cite{fong22}. The dashed colored lines represent the angular size that could potentially be covered by an example fibre bundle configuration, as shown in the sketch in the upper-right corner.}
    \label{petalo}
\end{figure}

However, this situation implies that more fibres should be used for each targeted galaxy in order to cover regions as large as the suitable projected offsets.
We refer to this type of fibres displaced in a group configuration as a 'fibre bundle' hereafter. This configuration is not currently included in the WST baseline but is under active consideration within the WST consortium. We emphasize that it would be extremely valuable for this multi-messenger science case.
In Figure \ref{petalo}, we show the distribution of the SGRB offsets from \cite{fong22,nugent22} for SGRB at $z<0.5$ and some representative fibre bundles so as to cover offsets up to $\sim5$\arcsec.

The cases where the EM counterpart explosion site would fall in the same fibre as the host galaxy centre would be especially difficult and require spectral subtraction of the host spectrum - ideally previously obtained with WST, thanks to WST survey.
We note that spectral subtraction can also be performed after the transient has faded, by acquiring the host galaxy spectrum within a few weeks of the event. While this approach would not allow detailed spectroscopic follow-up of the transient itself, it would still enable kilonova detection and redshift measurement of the host, having an impact on other science cases such as the determination of $H_0$.

Of course, the galaxies in the regions covered by IFS observations are less affected by these issues.

\begin{figure*}
\begin{center}
    \includegraphics[width = 0.45\linewidth]{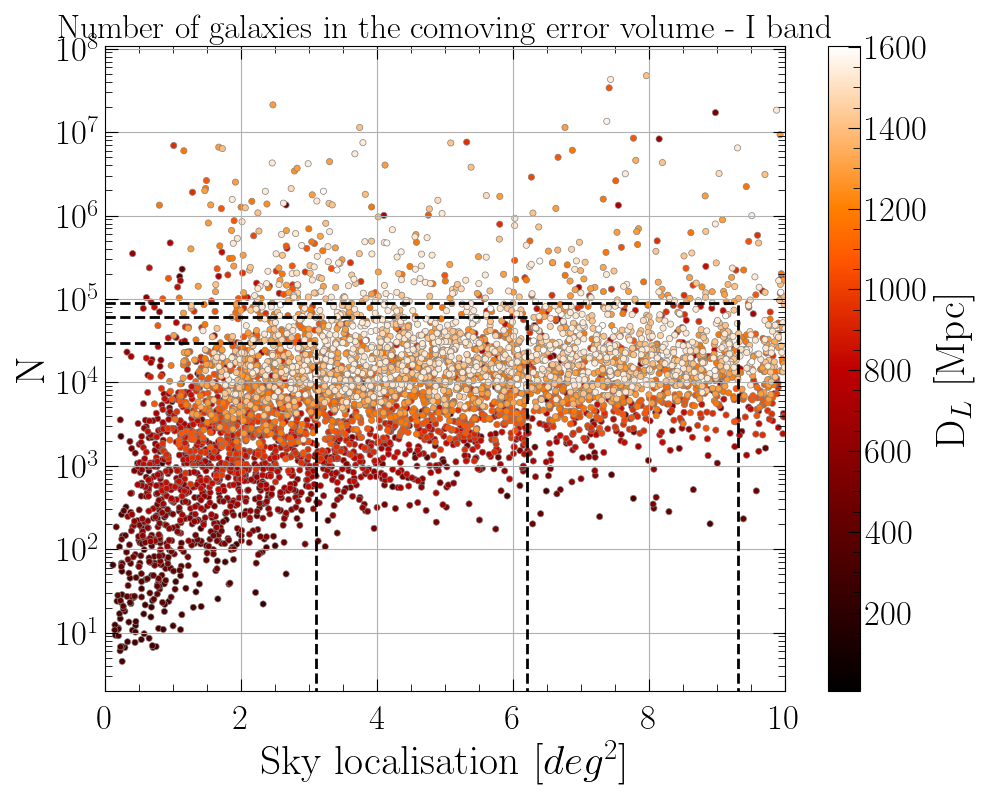}
    \includegraphics[width = 0.45\linewidth]{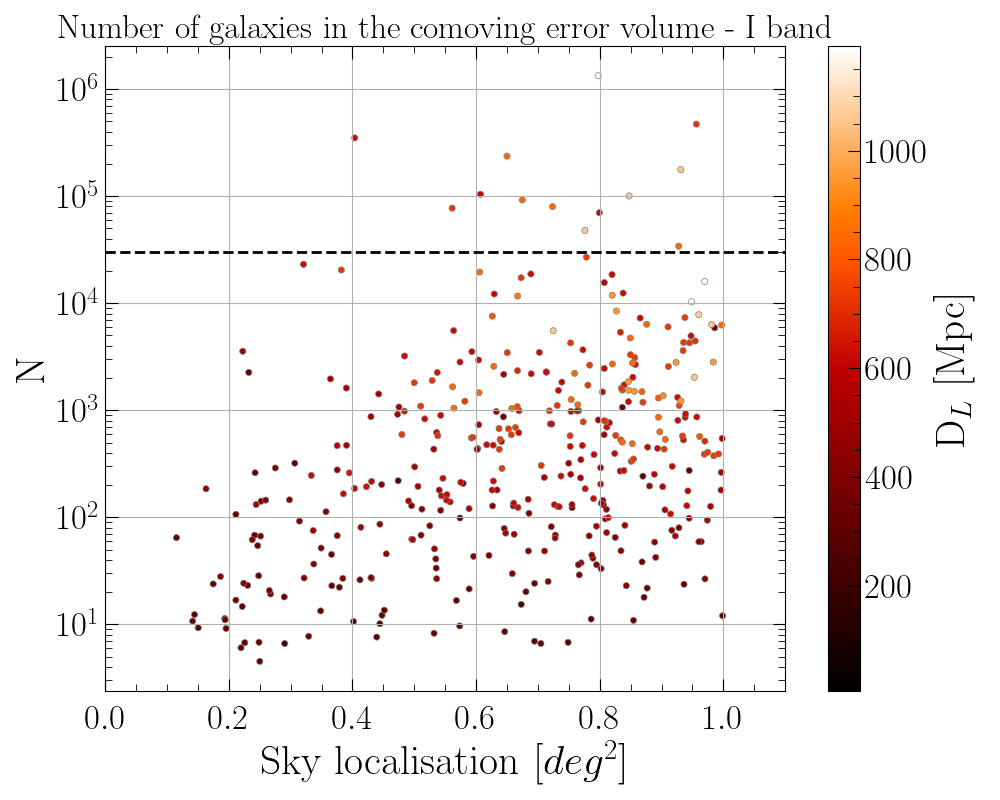}
    \caption{Number of galaxies in the comoving error volume of ET+CE BNS at $z<0.3$ with $\Omega<10\deg^{2}$ (left) and $\Omega<1\deg^{2}$ (right). The black dashed lines represent an indication of what WST can cover with three (left panel) and a single (right panel) one hour exposures, as described in section \ref{n_galaxies}.}
    \label{z0.2}
\end{center}
\end{figure*}

\subsection{The WST golden cases}\label{golden_cases}

From Fig. \ref{ETvsETCE_skyloc} and \ref{ilbert_I} we can define the golden cases for the WST standalone scenario.  
Up to redshift $ z \sim 0.3, 0.2$ for the ET+CE and ET alone configurations, respectively, it is possible to achieve sky localisations of the GW sources corresponding to a number of galaxies within the comoving error volume of BNS detections that can be easily covered with a few hours of WST observations. This is illustrated in Fig. \ref{z0.2}  
showing that, at $\Omega < 10\deg^{2}$, 3 one-hour WST exposures would cover the relevant number of galaxies for the vast majority of the events
(left panel of Fig. \ref{z0.2}). For the events with $\Omega<1\deg^{2}$ (right panel of Fig. \ref{z0.2}), we find that one exposure is sufficient for almost all the events.
By taking into account the use of fibre bundles, the same results are valid up to redshift $ z\sim0.2, 0.1$, for the ET+CE and ET alone configurations, respectively, when considering bundles of 20 fibres.

The events at $z<0.3$ (0.2) with a sky localisation $<10\deg^{2}$ would be more than 5000 (2000) over the whole sky within 10 years of observations with ET+CE and more than 100 with ET alone.

We stress that, even for the events with  small sky localisation regions, the comoving error volume can be large (See Fig. \ref{comoving_volume_skyloc}), and the role of WST multiplexing remains crucial for targeting the large number of galaxies within it.

As highlighted by \citeauthor{loffredo25}, significant improvements in sky localisation can also be achieved when the ET operates as part of a network with current detectors, LIGO, Virgo, KAGRA, and
LIGO-India at their design sensitivities. Within a redshift of 0.2, the number of events localised
within 10 square degrees is comparable to that obtained when ET is in a network with CE.

\subsection{Synergy with optical-NIR photometric observations}\label{syn}

WST can work in synergy with wide-field optical-NIR photometric facilities such as the Vera Rubin Observatory, assuming that it will be operational after the end of LSST. We note that, up to $z\sim0.2$, almost all KN that are detectable with Rubin are also detectable with both WST IFS and fibres (See Fig. \ref{z_distribution}).

In this framework, Rubin could systematically point the error regions of BNS detected by ET, performing deep observations and promptly identifying EM counterpart candidates. A substantial number of such candidates is expected to be distributed across the broad two-dimensional localisation regions, and spectroscopic data are the key to robustly identify and characterise the true counterpart. 

Based on extrapolations from the Zwicky Transient Facility \footnote{\url{https://www.ztf.caltech.edu/}} (ZTF) results, an average of around 200 new transients per $\deg^2$ per night from a single Rubin scan is expected (Julien Peloton, priv. comm.). This estimate refers to transient sources detected within a single visit, without any prior detection history, and includes objects that may later be classified as known phenomena, such as supernovae (SNe) or tidal disruption events (TDEs), through subsequent observations of the same area. 

In case of Rubin observations following a BNS detection, WST could target the transients detected by Rubin, by performing ToO observations to using just a fraction of the total number of fibres. Given the estimated density of $\sim$200 photometrically detected transients per $\deg^{2}$, approximately 600 fibres would be needed per pointing, considering WST field of view of $\sim3.1~\deg^{2}$. This would enable the rapid spectroscopic follow-up necessary for transient identification, while preserving the remaining fibre resources for other science cases.

In this scenario, the IFS would be advantageous for covering the high-density regions of newly discovered transients, coupled with the high-probability areas of the GW signal, and possibly also the regions where more extended galaxies are located.

We note that current transient rate estimates remain highly uncertain. However, the start of LSST operations at the Vera Rubin Observatory, expected before the end of 2025, will significantly improve our ability to constrain these rates and better inform future observing strategies.

\section{Summary and Conclusions}\label{conclusion}

In this work, we explored the potential of IFS and MOS capabilities of the WST for the detection, identification, and characterisation of EM counterparts to BNS detected by next-generation GW interferometers. 

To this purpose we used simulations of BNS merger populations, incorporating different neutron star equations of state and mass distributions. 
We simulated the corresponding GW signals considering both the $\Delta$ and $\mathrm{2L}$ ET designs, both for ET alone and in a network with CE. For each detected merger, we assigned a KN light curve based on AT2017gfo (\textit{gfo-like KN} sample) or numerical relativity-informed models (\textit{theoretical KN} sample). We also included simulations of GRB afterglow emission.

Rough spectra were constructed from the fit of the simulated photometric data and processed using WST ETC to compute the SNR in both WST IFS and MOS modes. Simulations show that WST can detect KNe up to redshift $z \sim 0.4 $ and magnitudes $ m_{\mathrm{AB}} \sim 25 $. GRB afterglows will primarily contribute for on-axis and slightly off-axis events, up to a viewing angle $ \Theta_{\mathrm{view}} \sim 15^\circ $, and can be detected up to higher redshifts (beyond $z=1$).

We discussed two main observing strategies: one with WST in a stand-alone scenario and the other in synergy with large field of view, sensitive photometric facilities, such as the Vera Rubin Observatory. 
In this last case, WST observations will likely use only a fraction of WST fibres, allowing simultaneous WST survey observations. 
Alternatively, in a standalone scenario, WST approximately 30\,000 fibres and the IFS can be employed in a galaxy-targeted strategy, covering the large number of galaxies in the wide GW error volumes. Improving the completeness of galaxy catalogues with redshift information at $z\le0.5$ will be essential for optimizing this strategy, enabling targeted observations of galaxies with redshifts consistent with the GW signal’s error volume. 

In our analysis, we considered the spatial offset between the EM counterparts and the centre of their host galaxies, as well as their respective brightness. We find that in most cases the EM counterparts will not lie beyond the effective radius of their host galaxies, and that the host galaxy surface brightness at the offsets of the EM counterpart would be comparable to or much fainter than the counterpart itself. Mini-IFUs with fibre bundles would be a solution to detect sources with large offsets from the host galaxy centre and allow for the coverage of extended galaxies at low redshift. Based on current offset distributions of KNe and SGRBs from their host centres, not using fibre bundles would likely result in detecting KNe in less than half of the cases. 
In some cases  the EM counterpart and its host galaxy may appear as superposed point sources. In such scenarios, spectral subtraction using previously acquired host galaxy spectra would be necessary. The acquisition of galaxy spectra up to $z\sim0.5$ as part of WST survey would be a great advantage to effectively perform this kind of subtraction.

Throughout this study, we limited our analysis to ET BNS with sky localisation uncertainties smaller than 100 $\deg^{2}$ (40 $\deg^{2}$ for ET in a network with CE). Even with these constraints, the expected number of events remains substantial, necessitating the development of additional prioritization criteria for optical follow-up. Those should first rely on information extracted from the GW signal, such as luminosity distance, to exclude events that are too distant to be detected and/or the inclination angle.

We identified a subset of $golden$ events for WST follow-up that are at redshifts $ z < 0.3 $ and that have sky localisations better than 10 $\deg^{2}$. These events are ideal
because they allow WST to cover all the galaxies in the error volume with a limited number of exposures. We estimate them to be from $\sim 10$ (ET-alone configuration) to hundreds per year (ET+CE configuration).   
WST contribution is particularly valuable also for well-localised events, because uncertainties in luminosity distance—even at low redshift—can still produce large three-dimensional error volumes containing numerous galaxies to target.

Regardless of whether WST operates in a standalone mode or in synergy with other observatories, we find that ToO observations for the research of KNe can begin within 12 to 24 hours after the GW detection. This changes if GRB prompt gamma-ray emission is detected by high-energy satellites at the same time of GW detection, suggesting an on-axis configuration. In such cases, WST becomes particularly valuable in case the GRB localisation is coarse ($\sim$arcminutes): the IFS can be efficiently used to scan the region, refine the position, and promptly identify the EM counterpart. Timely follow-up observations are especially critical for GRB afterglows, which follow rapidly decaying power-law lightcurves, with observations starting within a few hours after the GRB prompt detection.

In conclusion, the development of the future large field-of-view, high-sensitivity, IFS and high-multiplexing facilities should go beyond the more traditional science cases of galaxy survey and stellar population studies. The fact of handling the necessary depths, the large sky regions, and the vast numbers of objects to be targeted makes them a powerful way to 
address the challenges of the research of EM counterparts of next-generation GW-detected BNS mergers.  No existing MOS or IFS facility currently has the sensitivity and multiplexing capabilities required to fully address the BNS multi-messenger science case. The observational strategies are complex. They must be planned well in advance of operations, and take full advantage of the information available from the GW signal. Furthermore, to make the best scientific exploitation, such multi-messenger observations not only imply the capability to react on alerts with Target of Opportunity observations, but also to be able to perform a quasi-real-time quick data reduction and analysis so as to alert the community as soon as possible on the EM counterpart detection, in order to activate the most appropriate space- and ground-based facilities to perform deeper single-object observations to characterize the evolution of the EM counterpart and its properties.
It is clear that the multi-messenger science of the future will need new instruments having this science case included since the early phases of their development, and shaping their specifications and requirements.

\begin{acknowledgements}
The authors aknowledge Albino Perego, Andrew J. Levan,  Daniele Bjorn Malesani, Jan Harms and Julien Peloton for helpful discussion.
S.B. acknowledges support from the Erasmus+ programme.
S.B. and S.D.V. acknowledge support by the {\it API Gravitational waves and compact objects} of Paris Observatory and by the {\it Programme National des Hautes Energies (PNHE)} of CNRS/INSU co-funded by CNRS/IN2P3, CNRS/INP, CEA and CNES.
E.L. acknowledges funding by the European Union – NextGenerationEU RFF M4C2 1.1 PRIN 2022 project 2022RJLWHN URKA. 
M.B. and E.L. acknowledge financial support from the Italian Ministry of University and Research (MUR) for the PRIN grant METE under contract no. 2020KB33TP.
U.D. acknowledges Stefano Bagnasco, Federica Legger, Sara Vallero, and the INFN Computing Center of Turin for providing support and computational resources.
R.I.A. is funded by the Swiss National Science Foundation through an Eccellenza Professorial Fellowship (award PCEFP2\_194638).
The authors aknowledge the Astrophysics Centre for Multi-messenger studies in Europe (ACME) funded under the European Union's Horizon Europe Research and Innovation programme under Grant Agreement No 101131928.
The authors aknowledge the Wide-field Spectroscopic Telescope collaboration and the Research Group {\it Ondes Gravitationnelles}.
\end{acknowledgements}

   \bibliographystyle{aa} 
   \bibliography{bib} 

\pagestyle{empty}
\FloatBarrier
\begin{appendix}
\include{correctionblueband}
\include{exposuretime}

\include{additionalmaterial}

\end{appendix}

\end{document}

%% file: correctionblueband.tex
\section{Correction of the blue band for Set 1}\label{appendix2}

\begin{figure}
    \centering
    \includegraphics[width=\linewidth]{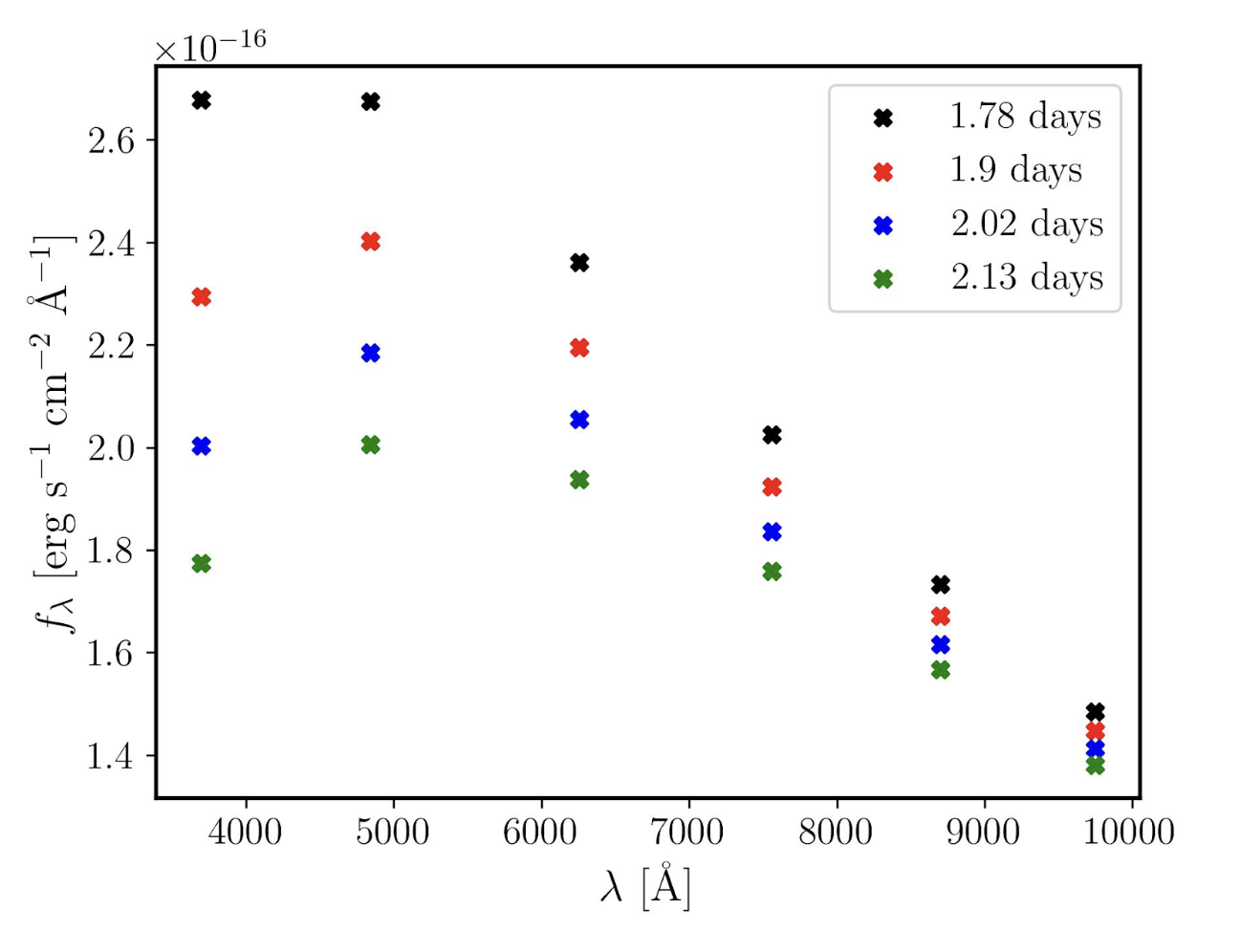}
    \caption{AT2017gfo-like KN f$_{\lambda}$ in Rubin filters as function of wavelength (observer frame), computed at four different time instants, assuming luminosity distance of 40 Mpc and viewing angle 25$^{\circ}$, in agreement with the value reported in column BF$_{\rm c}$ of Table 1 from \citealt{perego17} for AT2017gfo.}
    \label{corr_1}
\end{figure}

The limitations of the black-body approximation in modeling KN emission become evident in the UV band \citep{gillanders22}. KN ejecta, which are rich in heavy elements, exhibit strong line blanketing in the UV due to blue photons being absorbed in the opaque ejecta and re-emitted at redder wavelengths. The black-body approximation fails to capture this effect. Moreover, our KN modeling, based on \cite{perego17} and \cite{branchesi23}, assumes grey opacities, thereby neglecting spatial and temporal variations in opacity. This leads to an overestimation of the blue component in the spectrum. In Figure \ref{corr_1}, we present synthetic spectral continuum of AT2017gfo, computed from the light-curve model in \cite{perego17} used to construct Set 1 (see Sect. 2.1). A luminosity distance of 40 Mpc and a viewing angle of $\theta_{\mathrm{view}}$ = 25$^{\circ}$ were imposed, consistent with Table 1 in \cite{perego17}. By comparing Figure A.1 with AT2017gfo data (e.g., \cite{pian17}), it is evident that the synthetic spectra lack the pronounced spectral bump between approximately 4000$\AA$ and 5000$\AA$ observed in AT2017gfo. This discrepancy is primarily due to an overestimation of the blue component. Moreover, the bump in the synthetic spectrum emerges around 2 days, while the bump of AT2017gfo is at $\sim$1.44 days, suggesting a  $\pm 0.5$ day uncertainty in our model's bump timing.

To mitigate the overestimation of the UV component in our spectra, we devise a correction informed by the findings of \cite{gillanders22}. Their work used radiative transfer simulations to reproduce the spectra of AT2017gfo during photospheric emission, examining various ejecta compositions and classifying them based on the average electron fraction ($Y_e$). 
In their Figures 5 and 6, \citeauthor{gillanders22} compare the observed 1.4-day spectrum of AT2017gfo with model predictions for different $Y_e$ values, along with a \textit{continuum} spectrum. This continuum, calculated at the same temperature as the best-fit model ($Y_e$ = 0.37) but without accounting for ejecta composition, relies on the black-body approximation and, like our model, overpredicts the UV emission.

We extract fluxes corresponding to the \textit{continuum} and various $Y_e$ values ($Y_e = 0.37a, ~0.44a, ~0.05b$) from Figures 5 and 6 of \citet{gillanders22}, denoting them as $f_{\lambda, {\rm c}}$ and $f_{\lambda, Y_e}$, respectively. The \textit{continuum} and $Y_e = 0.37a$ fluxes are represented by the orange and blue lines in Figure 5, while the $Y_e = 0.44a$ and $Y_e = 0.05b$ fluxes correspond to the blue and purple lines in Figure 6, respectively. The difference between the black-body approximation and the radiative transfer simulations is expected to be influenced by the opacity of the ejecta \citep{tanaka20}. To quantify this discrepancy in the 3000 $\AA$ $\lesssim \lambda \lesssim $ 5000 $\AA$ region, we compute the following ratio for each $Y_e$:

\begin{equation}\label{eq:ratio_integrals}
   R = \frac{I_{Y_e}}{I_{\rm c}}~, 
\end{equation}
where $I_{\rm c}$ and $I_{Y_e}$ are the integrals of the \textit{continuum} and $Y_e$ fluxes over $\lambda$:
\begin{equation}\label{eq:flux_integrals}
    I = \int_{\lambda_\text{min}}^{\lambda_\text{max}} f_{\lambda} \, \mathrm{d}\lambda~.
\end{equation}

\begin{figure}
    \centering
    \includegraphics[width=\linewidth]{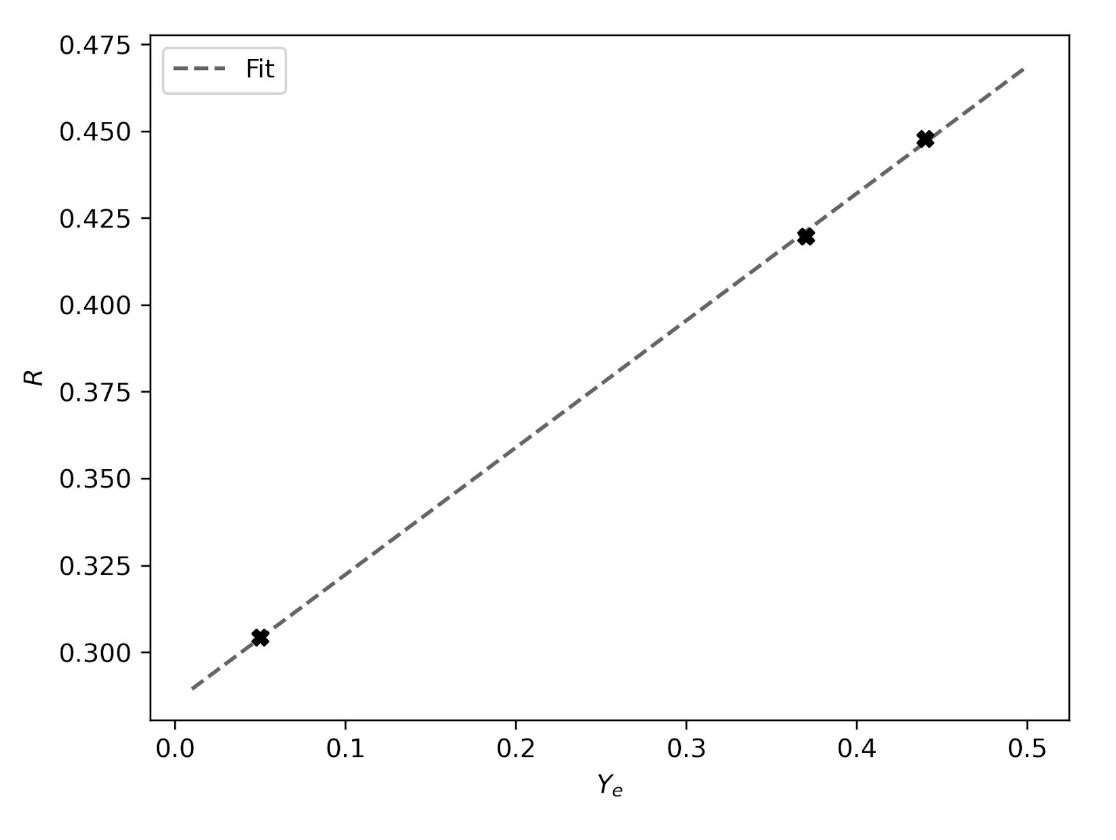}
    \caption{Ratio of flux integrals $R = I_{\rm c}/I_{Y_e}$ as defined in Equation \ref{eq:ratio_integrals}, calculated using the fluxes extracted from Figures 5 and 6 of \citealt{gillanders22}. These fluxes correspond to the \textit{continuum} (orange line in Figure 5), $Y_e = 0.37a$ (blue line in Figure 5), $Y_e = 0.44 a$ (blue line in Figure 6), and $Y_e = 0.05 b$ (purple line in Figure 6). The dashed grey line represents the linear fit as detailed in Eq. \ref{eq:fit_correction}}.
    \label{corr_2}
\end{figure}

We set $\lambda_{\rm min} = 3000$ $\AA$. The upper bound $\lambda_{\rm max}$ depends on $Y_e$: for lanthanide-poor ejecta ($Y_e > 0.3$), we use $\lambda_{\rm max} = 4250$ $\AA$, while for lanthanide-rich ejecta ($Y_e \leq 0.3$), we use $\lambda_{\rm max} = 5000\AA$. Figure \ref{corr_2} shows the computed ratio $R$ as a function of $Y_e$ (black crosses). We observe a linear correlation between $R$ and $Y_e$, and a linear fit yields:

\begin{equation}\label{eq:fit_correction}
    R = A + B~Y_e~,
\end{equation}
where $A = 0.2857~{\rm and}~B = 0.3658$, as indicated by the grey dashed line in Figure \ref{corr_2}. The ratio $R$ serves as a correction factor for adjusting the KN fluxes from \cite{branchesi23}.

To implement the correction to the KN fluxes, we first evaluate the KN flux, $f_{\lambda_{\rm s}}(t_{\rm s})$, in the source frame, where $\lambda_{\rm s}$ and $t_{\rm s}$ represent the emission wavelength and time, respectively. For times between 0.5 and 3 days, when the emission is primarily dominated by the dynamical ejecta, we use the dynamical opacity k$_{\rm dyn}$ to infer the corresponding electron fraction Y$_e$, based on Table 1 of \citet{tanaka20}. In our model of AT2017gfo-like spectra, k$_{\rm dyn}$ varies with the polar viewing angle \citep{perego17}. For viewing angles $\theta_{\rm v} < \pi/6$, we assume k$_{\rm dyn} = 1~\rm{cm \cdot g^{-1}}$, yielding Y$_e = 0.4$; for larger angles, we use k$_{\rm dyn} = 30~\rm{cm \cdot g^{-1}}$, corresponding to Y$_e = 0.1$.

For t$_{\rm s} \geq 3$ days, when the emission is dominated by the secular ejecta, we adopt a secular opacity of k$_{\rm s} = 5~\rm{cm \cdot g^{-1}}$, from which we infer Y$_e = 0.31$, again using Table 1 of \citet{tanaka20}.

Once the appropriate value of Y$_e$ has been determined, we compute the correction factor R using Equation~\ref{eq:fit_correction} and apply it to the flux in the source frame. The corrected flux is given by

 \begin{equation}\label{eq:applied_correction}
        f_{{\rm corr},\lambda_{\rm s}} = R \cdot f_{\lambda_{\rm s}}~.
 \end{equation}
 
We apply this correction up to $\lambda_{\rm s} = 4250$ $\AA$~for $Y_e > 0.3$, and $\lambda_{\rm s} = 5000$ $\AA$~otherwise.

\begin{figure}
    \centering
    \includegraphics[width=\linewidth]{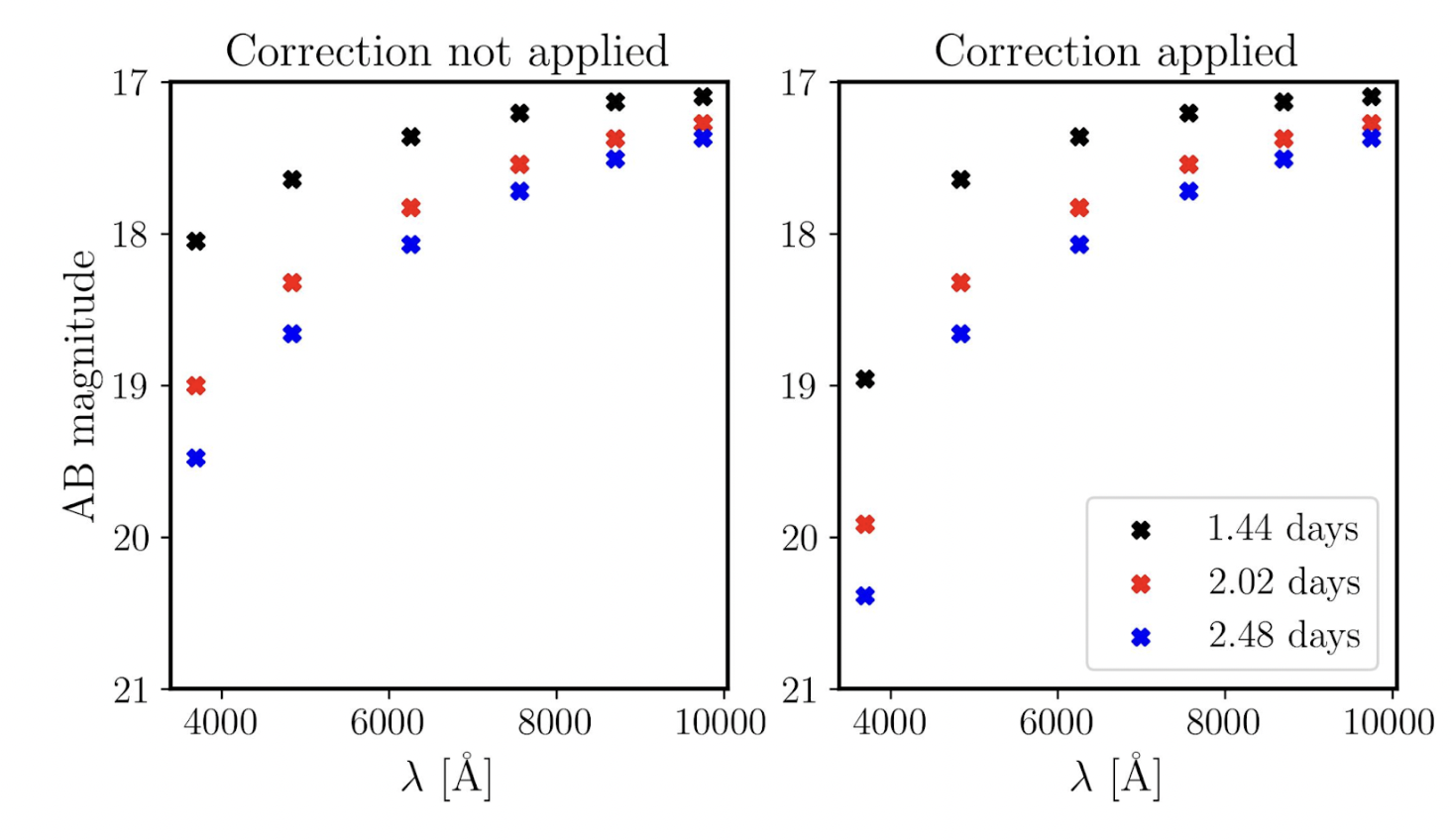}
    \caption{Our observer frame spectrum of an AT2017gfo-like KN, computed at three distinct time points with an assumed luminosity distance of 40 Mpc and a viewing angle of 25 degrees. The left panel displays the spectrum before applying any correction, while the right panel illustrates the spectrum after applying the flux correction as detailed in Equation \ref{eq:applied_correction}. A comparison with the AT2017gfo observational data, see for example \citealt{villar17}, reveals that the u filter emission (3694 \AA) is initially too high. After implementing the correction, the emission aligns more closely with the observational data.}
    \label{corr_3}
\end{figure}

In Figure \ref{corr_3}, we show the synthetic spectrum computed before and after applying the correction of Eq. \ref{eq:applied_correction} for an AT2017gfo-like event, located at the luminosity distance of 40 Mpc and observed with a 25$^{\circ}$ viewing angle.
Comparing our synthetic spectrum both before and after applying the correction with AT2017gfo data, e.g., Figure 4 of \cite{villar17}, we find that the emission in the u filter ($3694\, \AA$) is excessively high before applying the correction (approximately 19 magnitudes two days after the merger), but aligns more closely with the AT2017gfo observation after the correction (approximately 20 magnitudes two days after the merger).\\

Our method for generating KN spectral continuum is an approximation, mainly because of the limited frequency range, the employment of grey opacities, and the assumption of black-body emission. Even with these limitations, our estimated population of KN spectral curves remains valuable, serving as a critical tool in calibrating the features of next-generation spectroscopic telescopes and evaluating their performance.

%% file: exposuretime.tex
\section{Exposure Time}\label{appendix}
We present in this Appendix the same results reported in section \ref{results}, but considering this time 1800s exposures instead of 3600\,s ones.

\begin{figure}
    \centering
    \includegraphics[width = 0.49\linewidth]{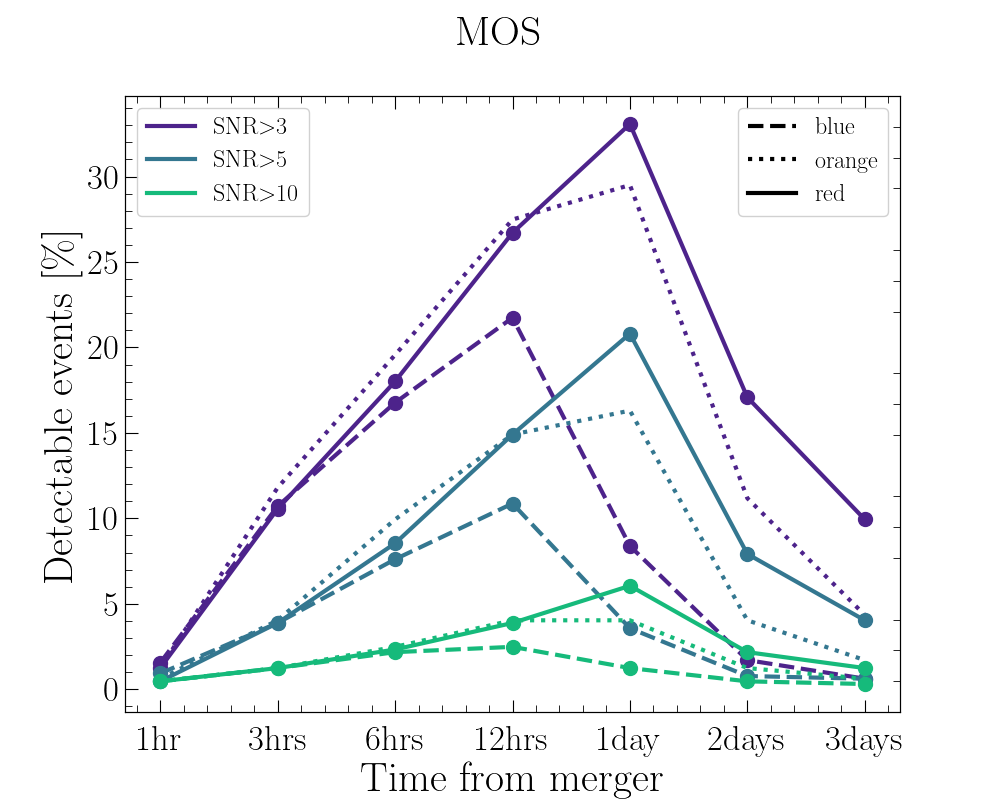}
\includegraphics[width = 0.49\linewidth]{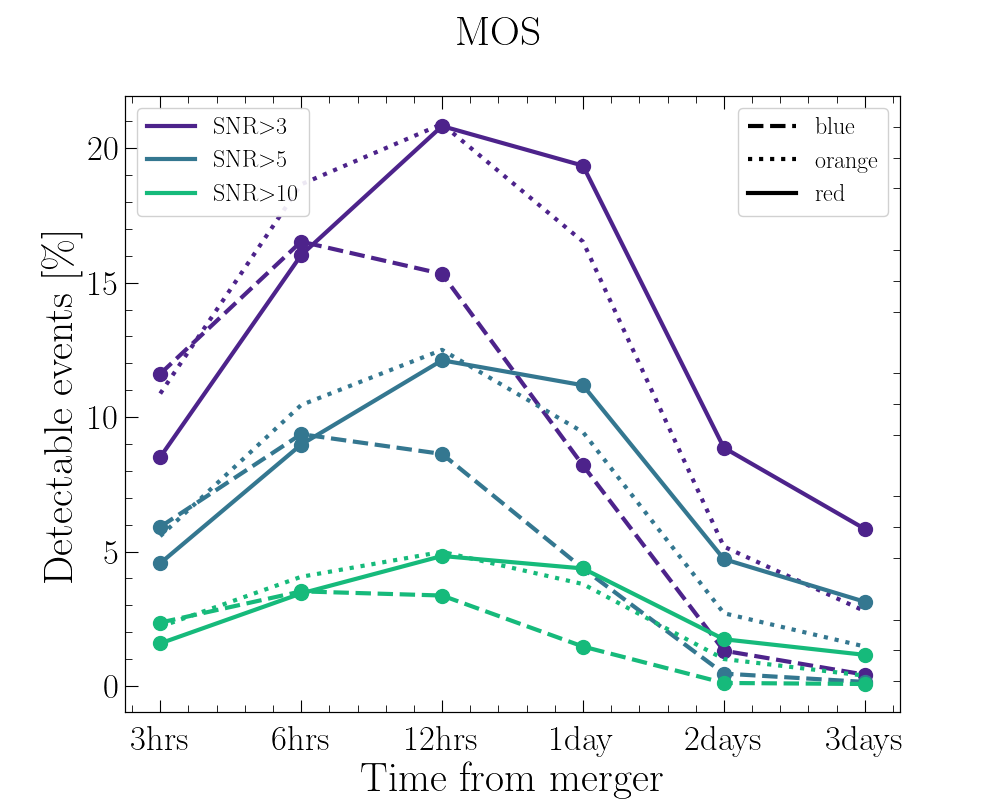}
    \caption{}
    \label{fig:enter-label}
\end{figure}

\begin{figure}
    \centering
    \includegraphics[width=\linewidth]{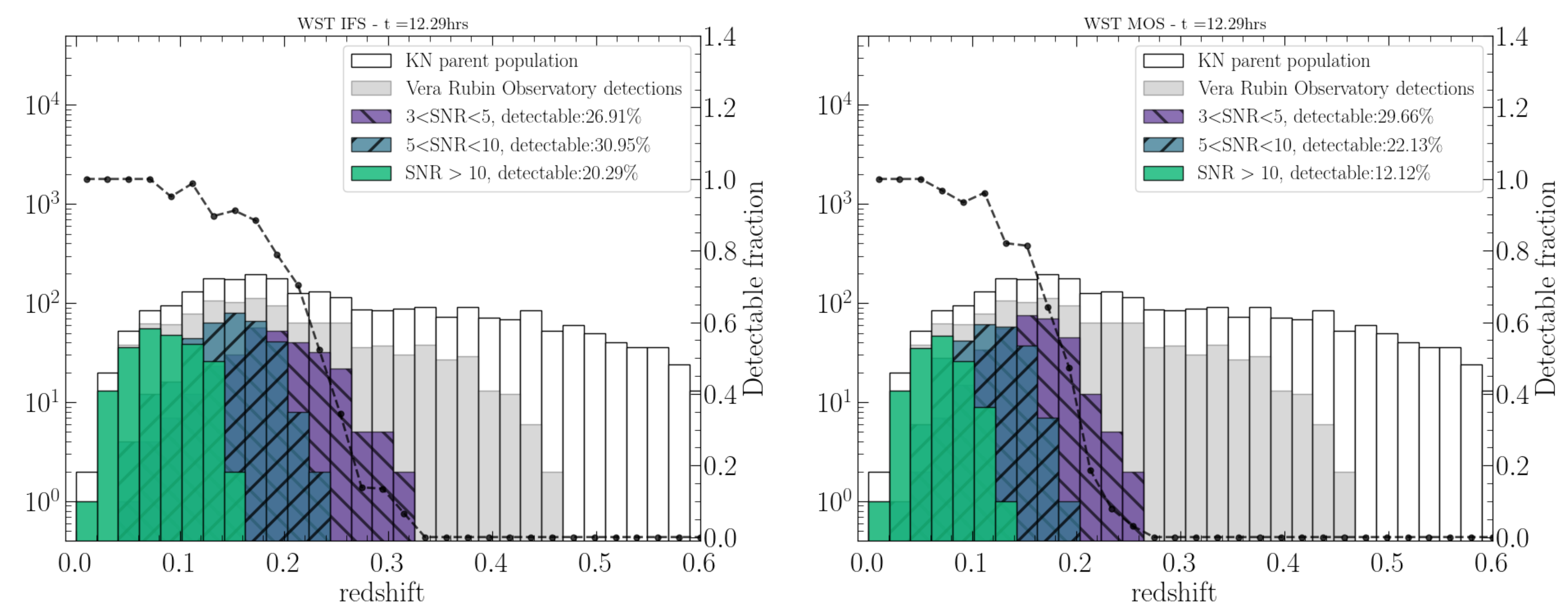}
    \caption{}
    \label{fig:enter-label}
\end{figure}

\begin{figure}
    \centering
    \includegraphics[width=0.49\linewidth]{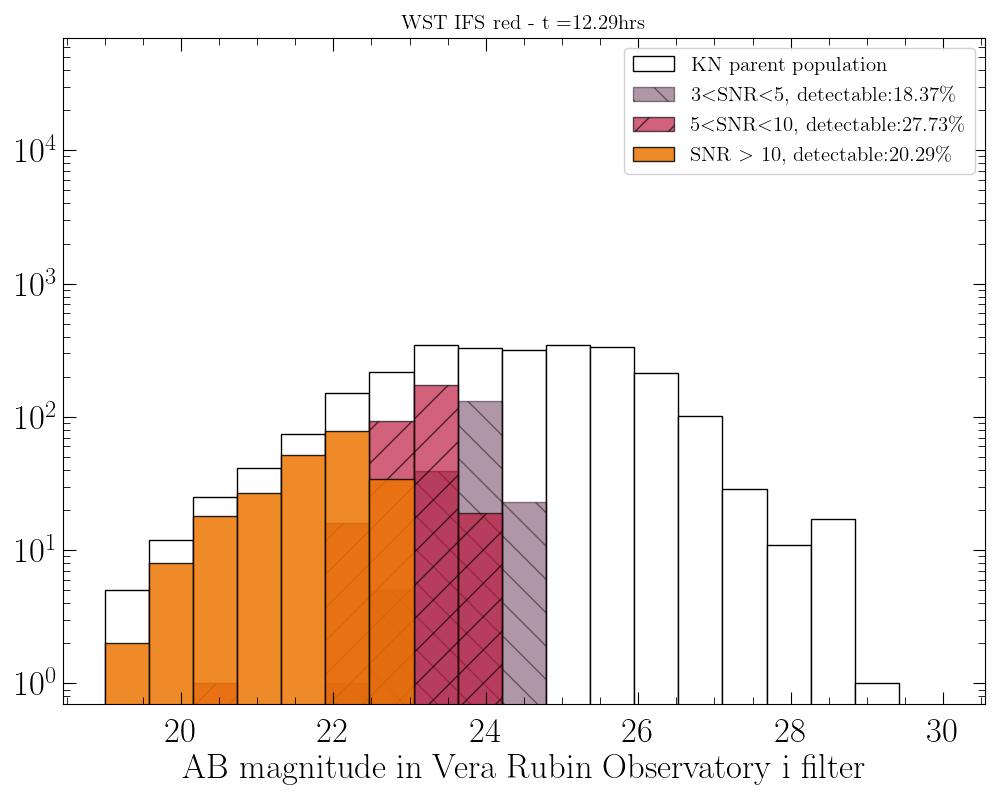}
    \includegraphics[width=0.49\linewidth]{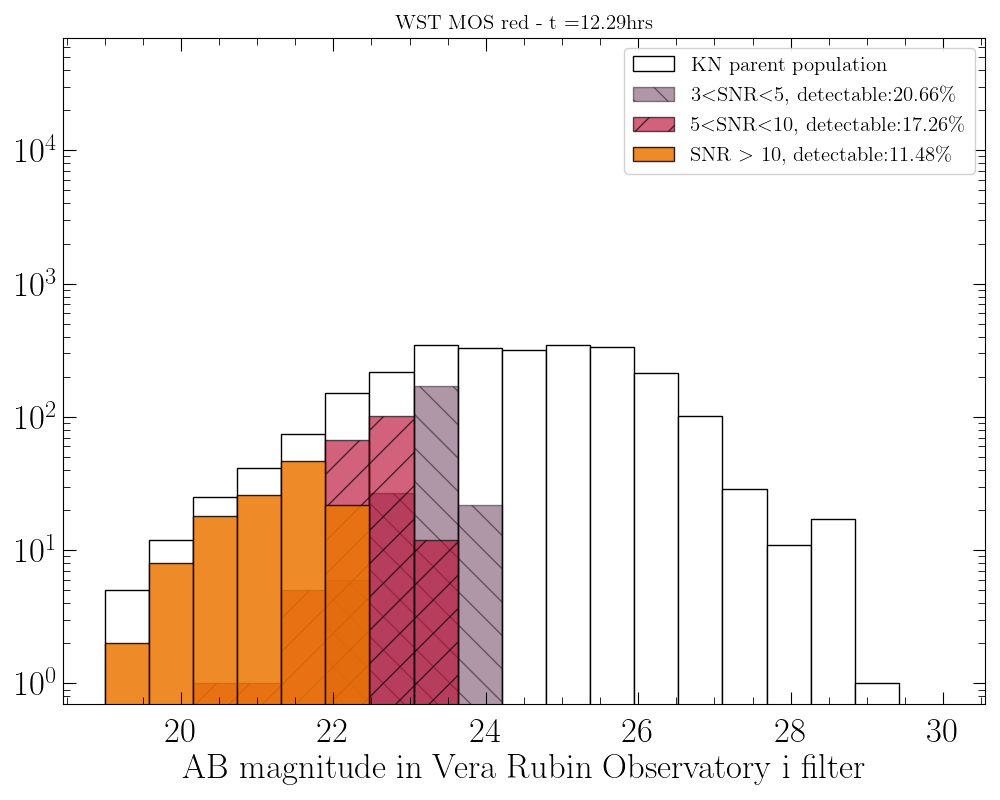}
    \caption{}
    \label{fig:enter-label}
\end{figure}

\begin{figure}
    \centering
    \includegraphics[width=0.49\linewidth]{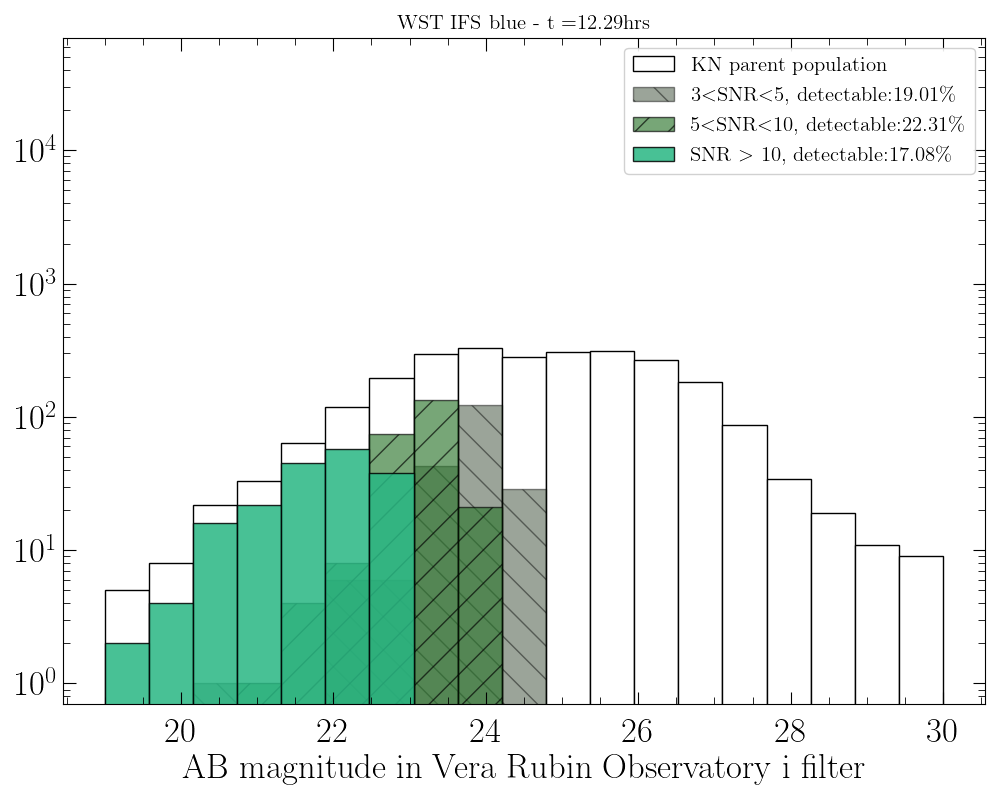}
    \includegraphics[width=0.49\linewidth]{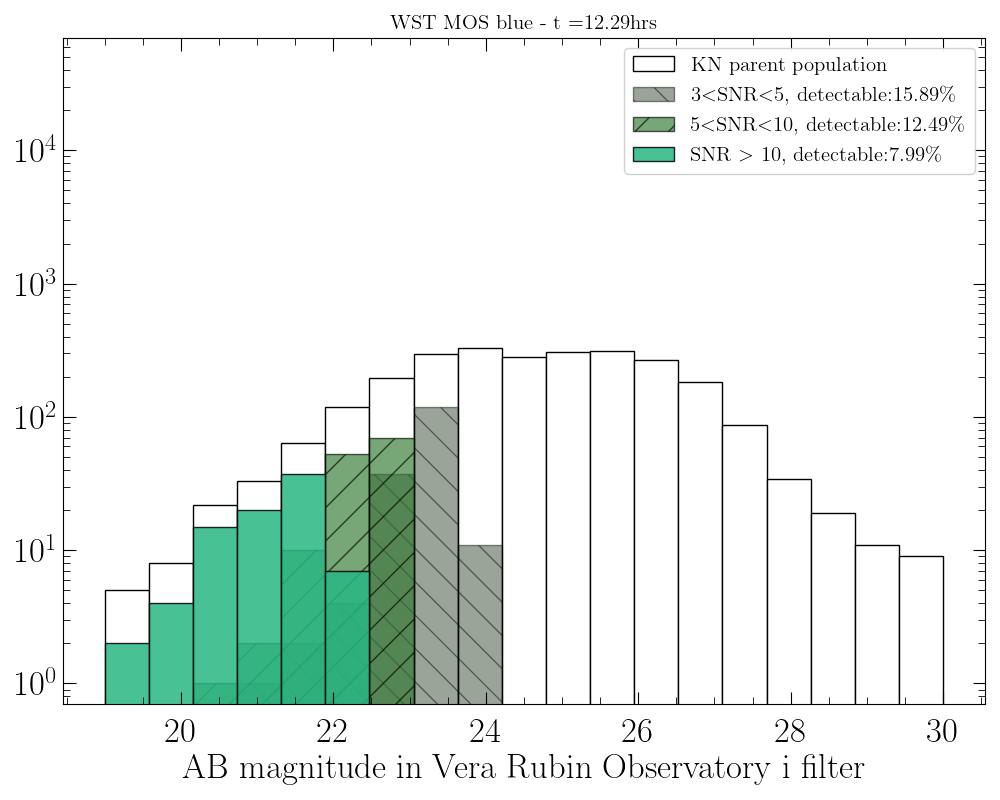}
    \caption{}
    \label{fig:enter-label}
\end{figure}

\begin{figure}
    \centering
    \includegraphics[width=0.49\linewidth]{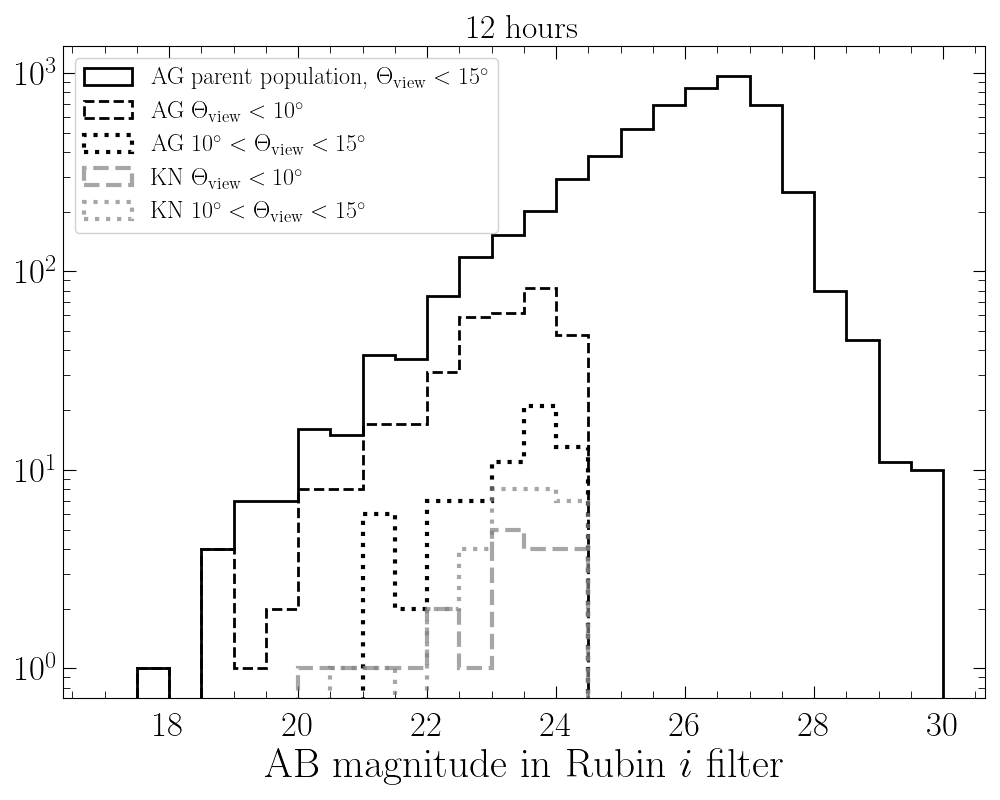}
    \includegraphics[width=0.49\linewidth]{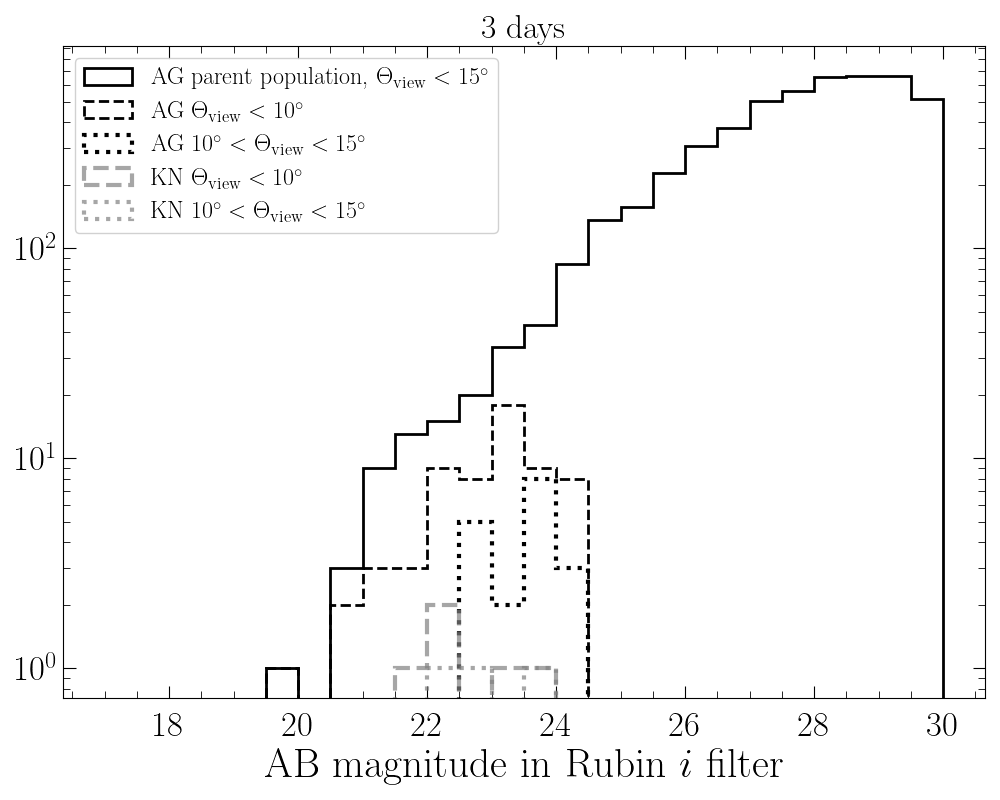}
    \caption{}
    \label{fig:enter-label}
\end{figure}

%% file: additionalmaterial.tex
\section{Additional material}\label{appendix3}

\begin{figure}[h!]
\centering
\includegraphics[width = \linewidth]{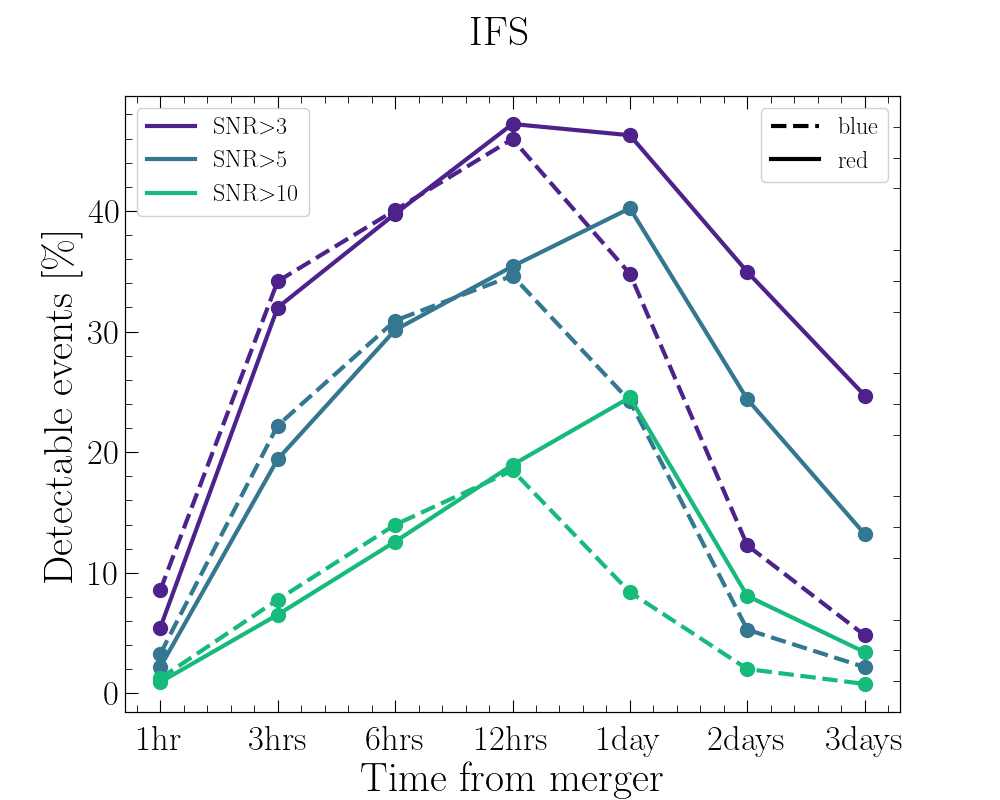}
\includegraphics[width = \linewidth]{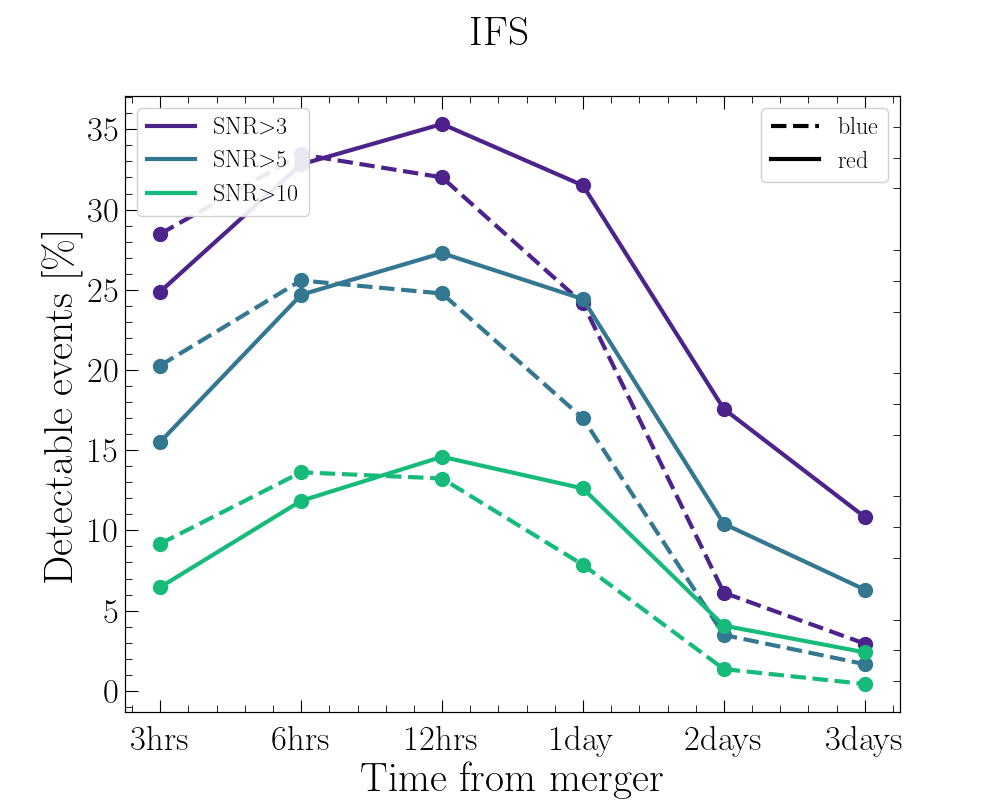}
\caption{Percentage of \textit{gfo-like} KNe (top) and \textit{theoretical} KNe (bottom) following BNS mergers detected by ET in the 2L configuration that are detectable with WST IFS.
Different colors indicate different SNRs; dashed and solid lines correspond to the blue and red arms of the spectrographs, respectively.
The highest detectability occurs approximately 12--24 hours post-merger.
}
\label{detectability_2L_IFS}
\end{figure}

\begin{figure}
\centering
\includegraphics[width = \linewidth]{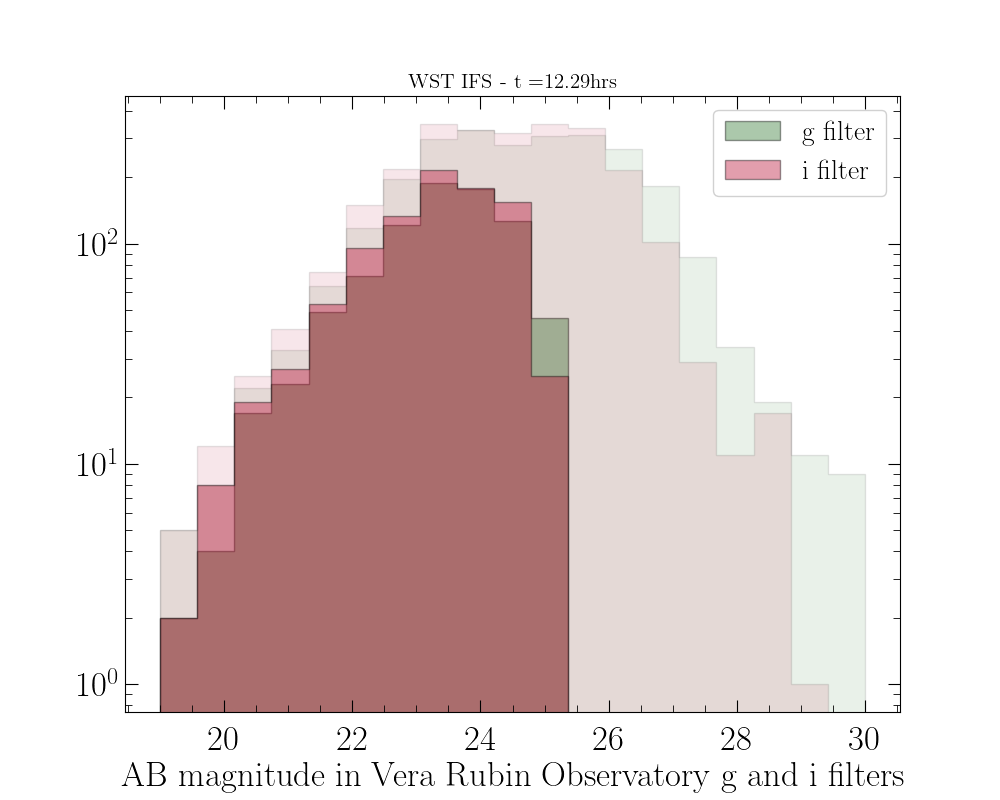}
\includegraphics[width = \linewidth]{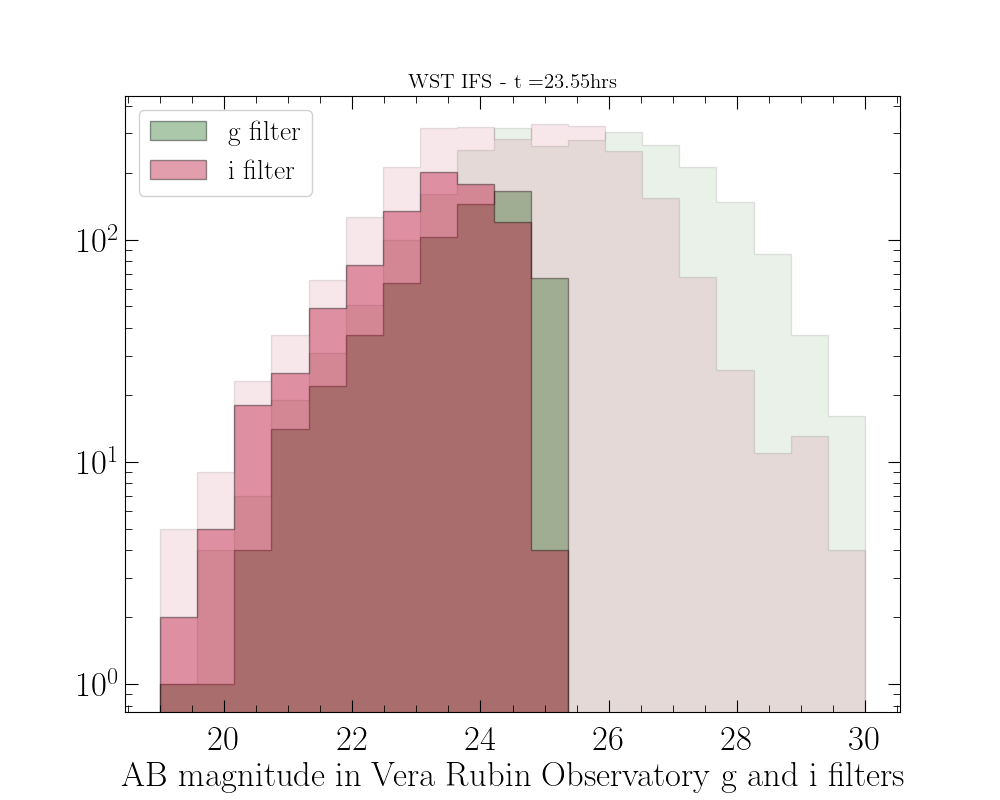}
\includegraphics[width = \linewidth]{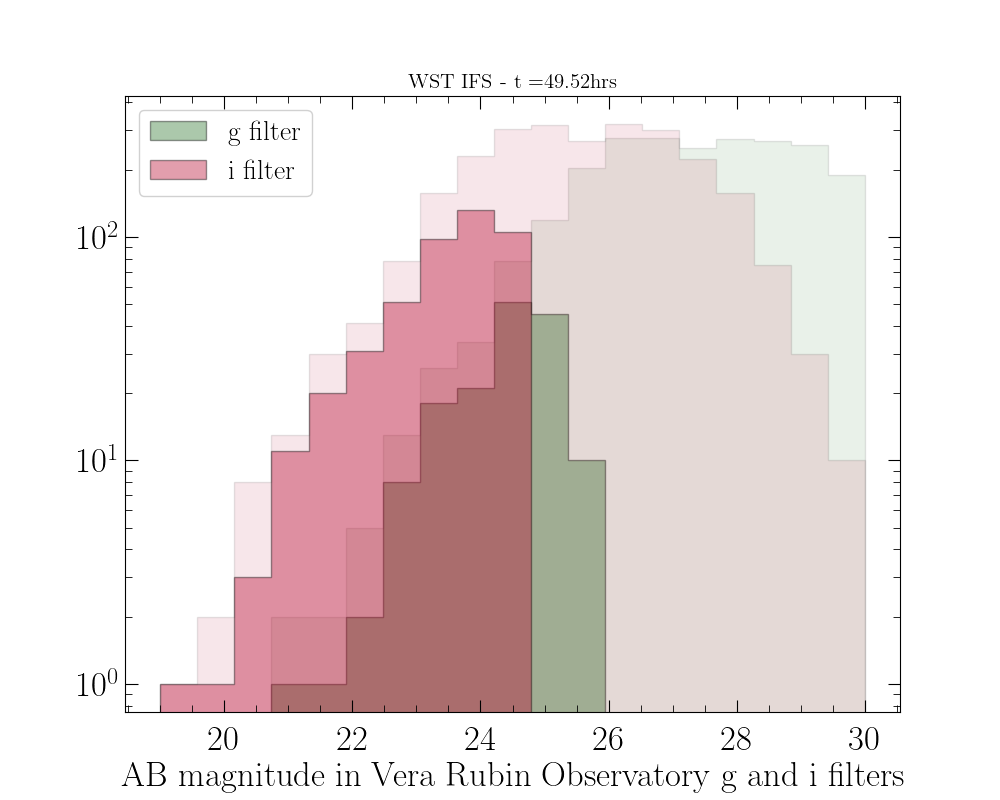}
\caption{Magnitude distribution in $i$ and $g$ filters of KNe that are detectable with WST at $\sim$ 12 (left), 24 (center) and 48 (right) hours post-merger. The background histograms show the BNS+KN parent population.}
\label{mag_i_vs_mag_g}
\end{figure}

\begin{figure}
\centering
\includegraphics[width = \linewidth]{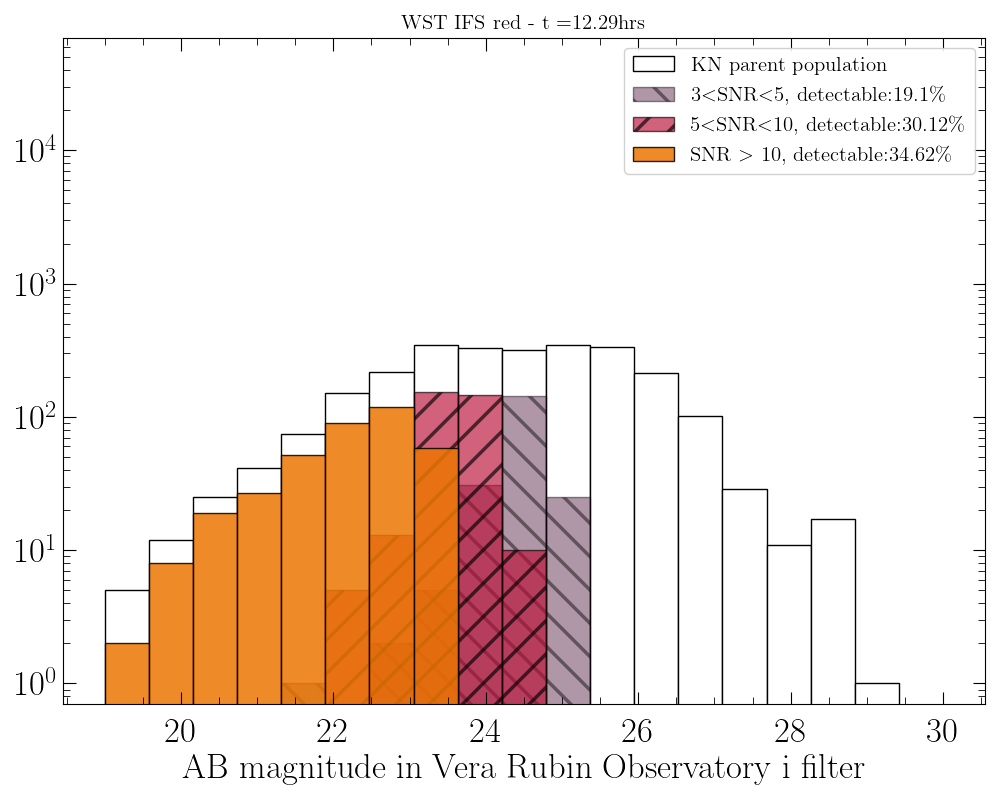}
\caption{Magnitude distribution in $i$ filter of ET BNS detected over 10 years and their corresponding \textit{theoretical KN} at $\sim$ 12 hours after the merger.  
The background distribution in white corresponds to the parent BNS+KN population. 
The colored distributions correspond to the EM counterparts that are detectable with WST IFS red arm, different colors corresponding to different SNR intervals.}
\label{mag_i_distribution_IFS}
\end{figure}

\begin{figure}
\centering
\includegraphics[width = \linewidth]{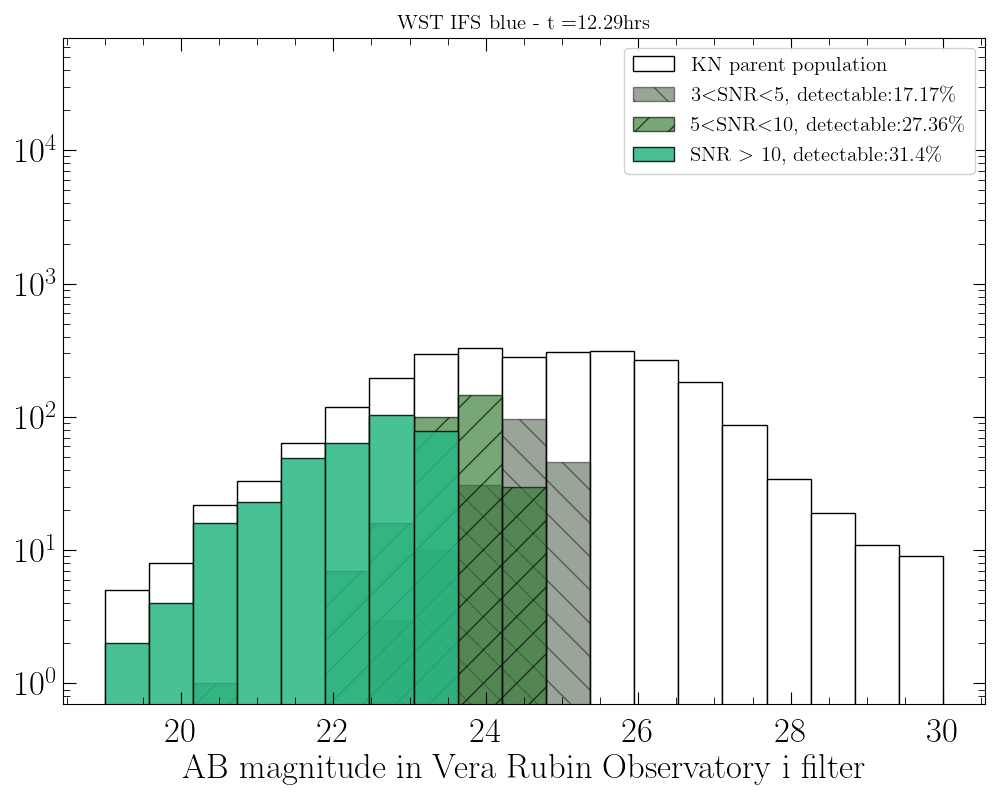}
\includegraphics[width = \linewidth]{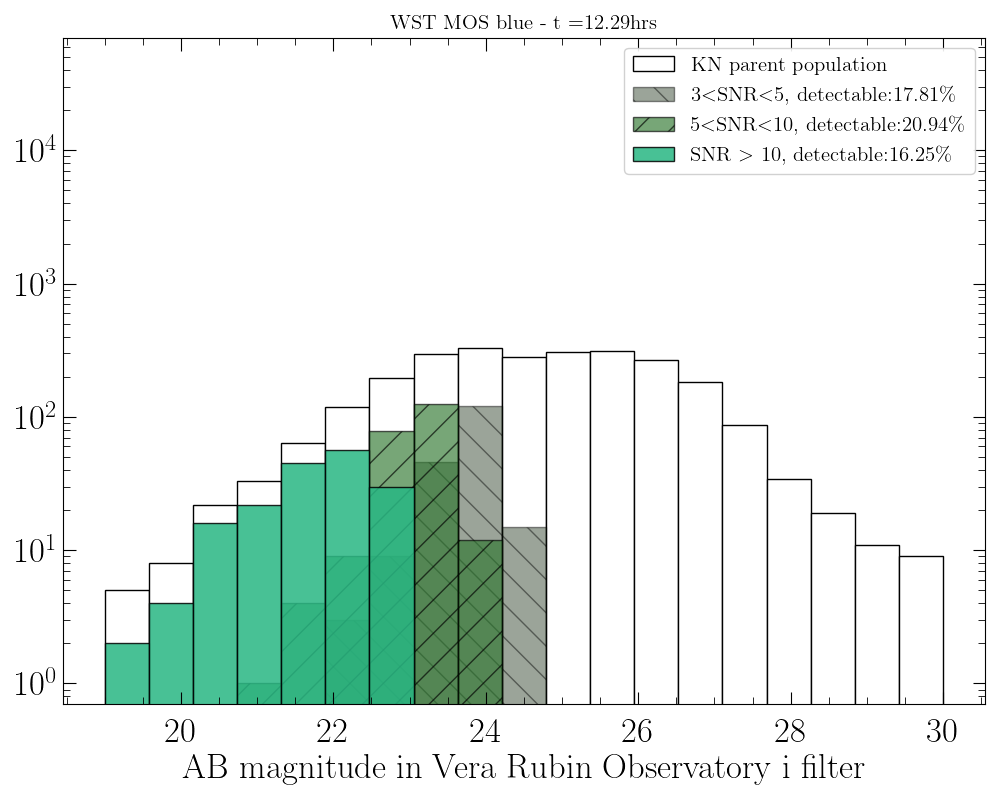}
\caption{Magnitude distribution in $g$ filter of ET BNS and corresponding theoretical KN at $\sim$ 12 hours after the merger.
The background distribution in white corresponds to the parent BNS+KN population. 
The colored distributions correspond to the EM counterparts that are detectable with WST IFS (top panel) and MOS (bottom panel) blue arm, different colors corresponding to different SNR intervals. EM counterparts detectable with Rubin are shown in grey.}
\label{mag_g_distribution}
\end{figure}

\begin{figure}[h!]
\centering
\includegraphics[width = \linewidth]{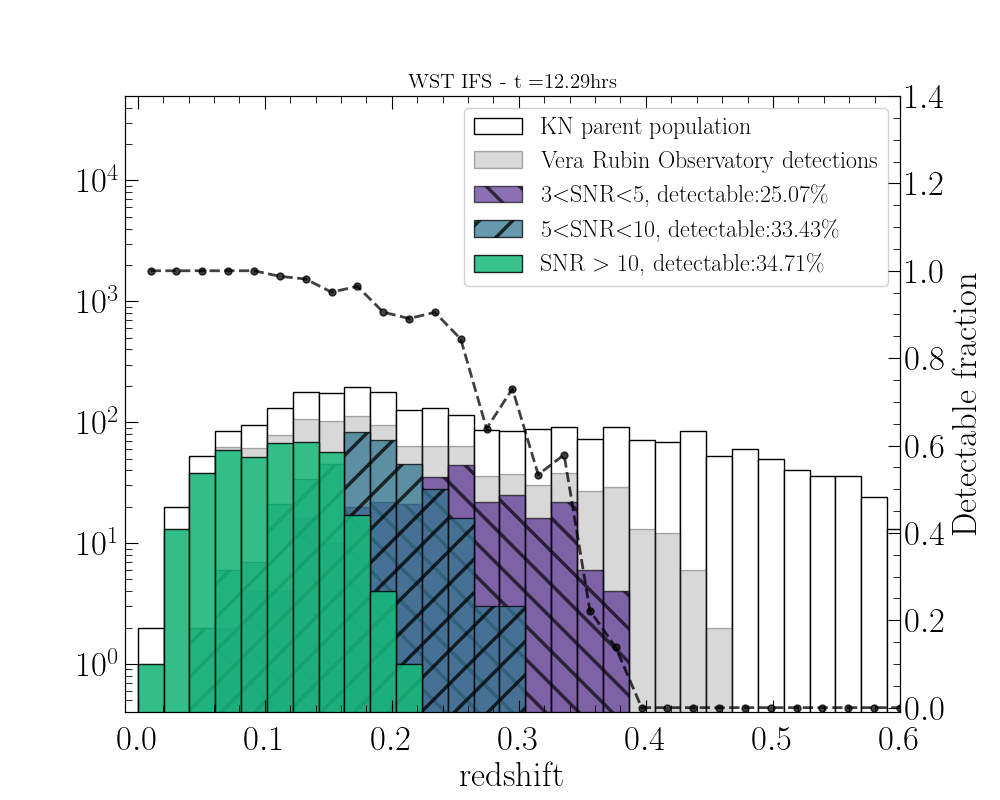}
\caption{Redshift distribution of ET BNS detected over 10 years and the corresponding \textit{theoretical KN} at $\sim$ 12 hours after the merger.
The background distribution in white corresponds to the parent BNS+KN population. 
The colored distributions correspond to the EM counterparts that are detectable with WST IFS blue or red arm, different colors corresponding to different SNR intervals. EM counterparts detectable with Rubin are shown in grey. Black points refer to the y axis ticks on the right hand side and show the fraction of detectable (SNR $>$ 3) EM counterparts, with no additional information on the SNR range, with respect to Rubin detections.}
\label{z_distribution_IFS}
\end{figure}

\begin{figure}
    \centering
    \includegraphics[width=0.9\linewidth]{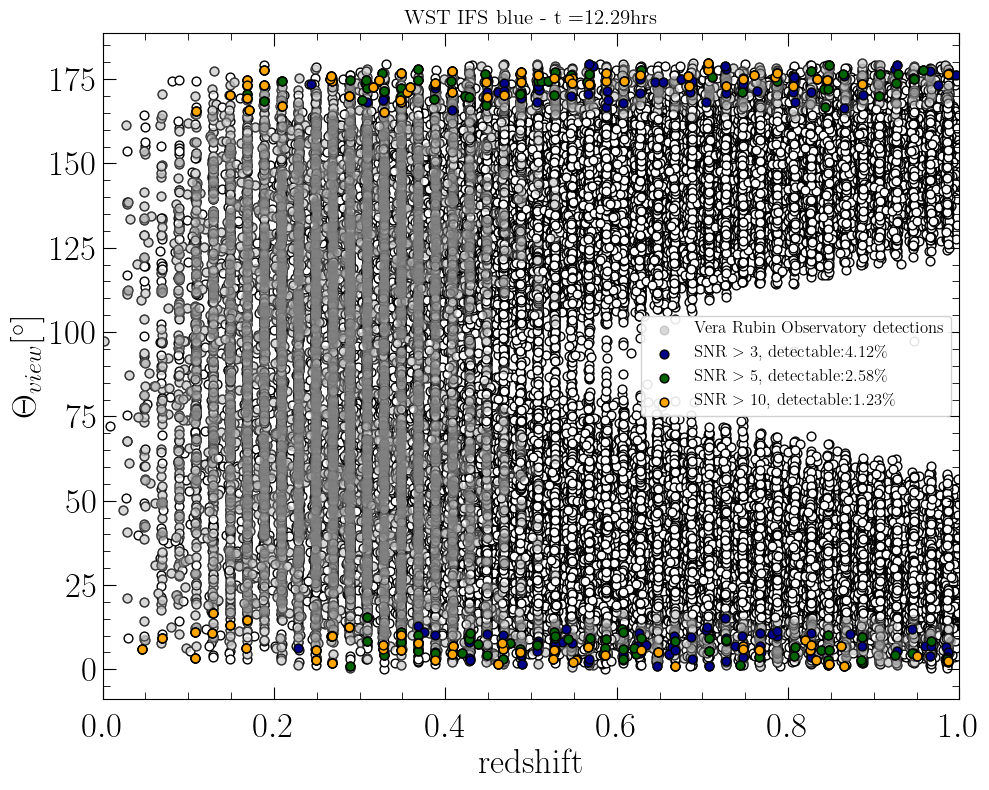}
    \caption{Viewing angle as a function of redshift of GRB afterglows following BNS mergers (white points) detected by ET in a network with CE. Colored points represent WST detections, different colors corresponding to different SNR. Percentages are given with respect to the number of Rubin detections.}
    \label{theta_agkn}
\end{figure}

\begin{figure}
    \centering
    \includegraphics[width = \linewidth]{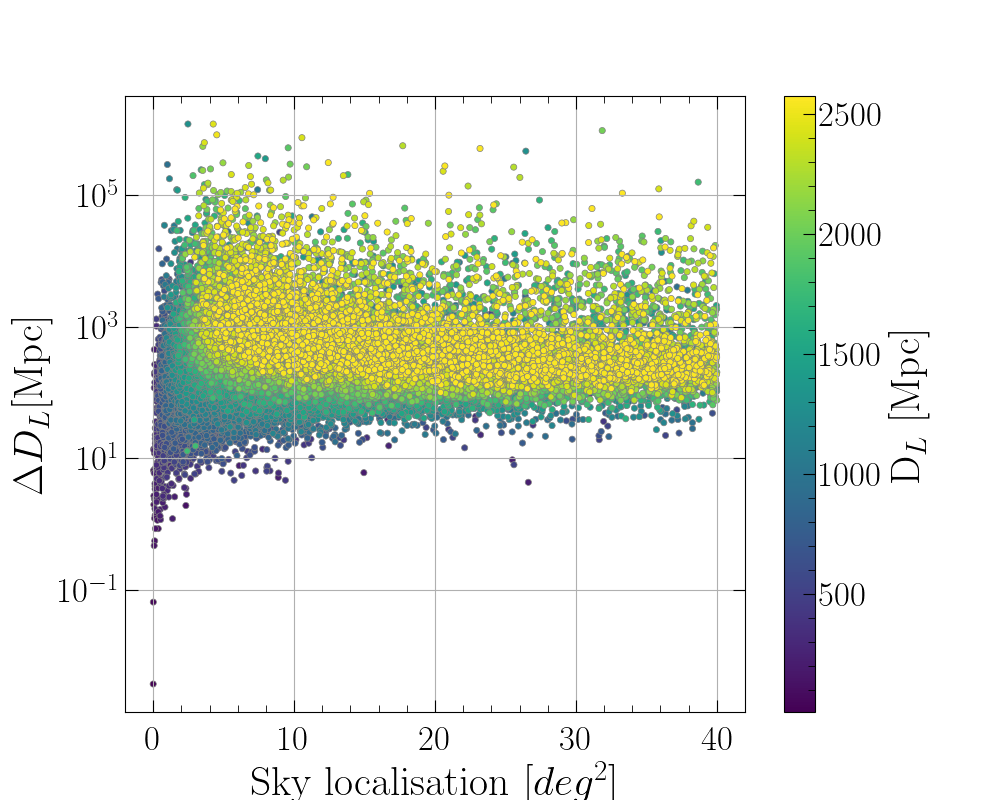}
    \caption{Uncertainty on the GW luminosity distance as a function of sky localisation for each BNS detected by ET in a network with CE, at $z<0.45$. The color of the points is related to the luminosity distance D$_{L}$ of the events as reported in the lateral color bar.}
    \label{comoving_volume_skyloc}
\end{figure}

\begin{figure}
    \centering
    \includegraphics[width=\linewidth]{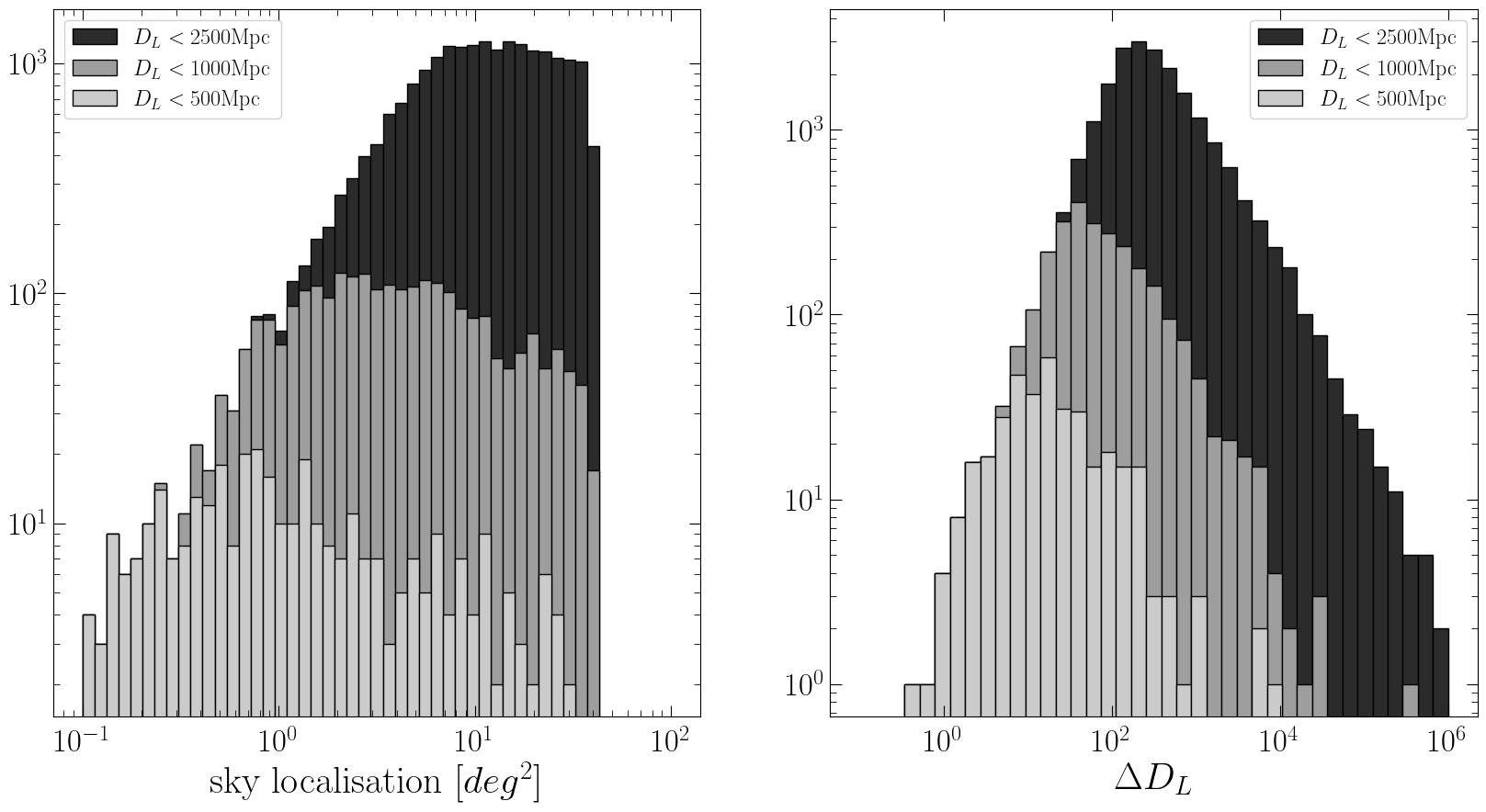}
    \caption{Distribution of sky localization (left panel) and uncertainty on $D_{\mathrm{L}}$ (right panel) for different values of $D_{\mathrm{L}}$ of BNS detected by ET in a network with CE.}
    \label{skylocDL}
\end{figure}

\begin{figure}
    \centering
    \includegraphics[width = \linewidth]{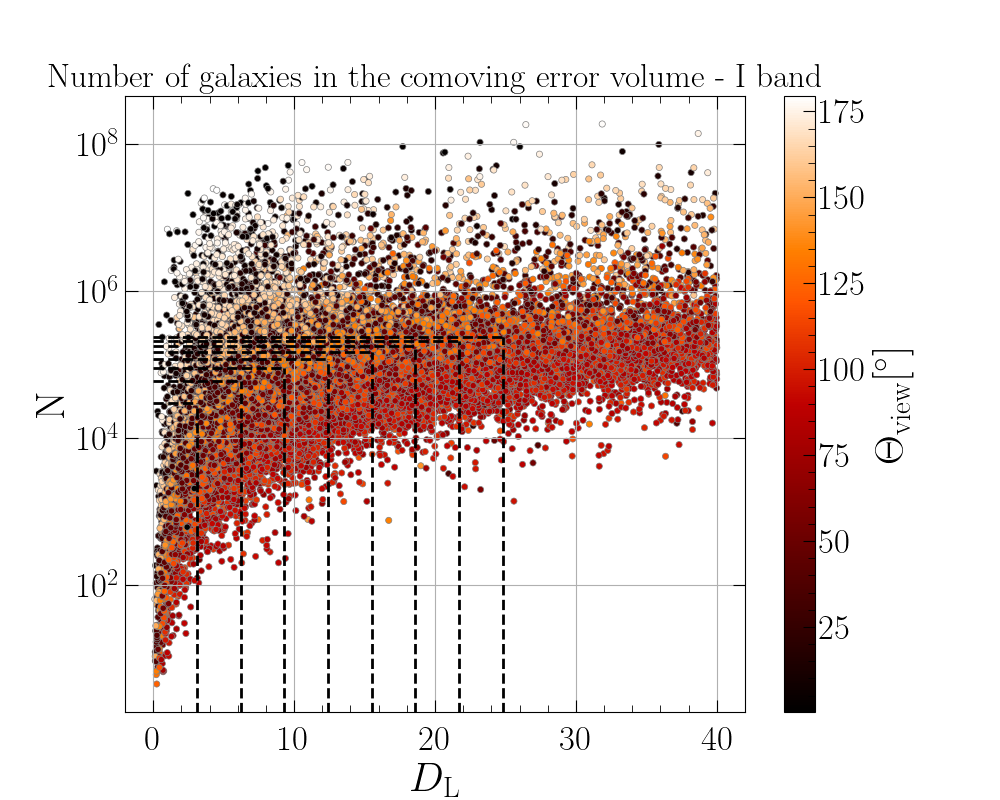}
    \caption{Number of galaxies in the comoving error volume of BNS at $z<0.45$ detected by ET in a network with CE, as a function of their sky localisation. Vertical dashed lines represent the sky area covered for increasing number of exposures. Horizontal dashed lines refer to the corresponding number of WST fibers available within the FoV considered. The color bar refers to the viewing angle $\Theta_{\rm view}$.}
    \label{ilbert_I_skyloc_ETCE_coltheta}
\end{figure}

\begin{figure}
    \centering
    \includegraphics[width = \linewidth]{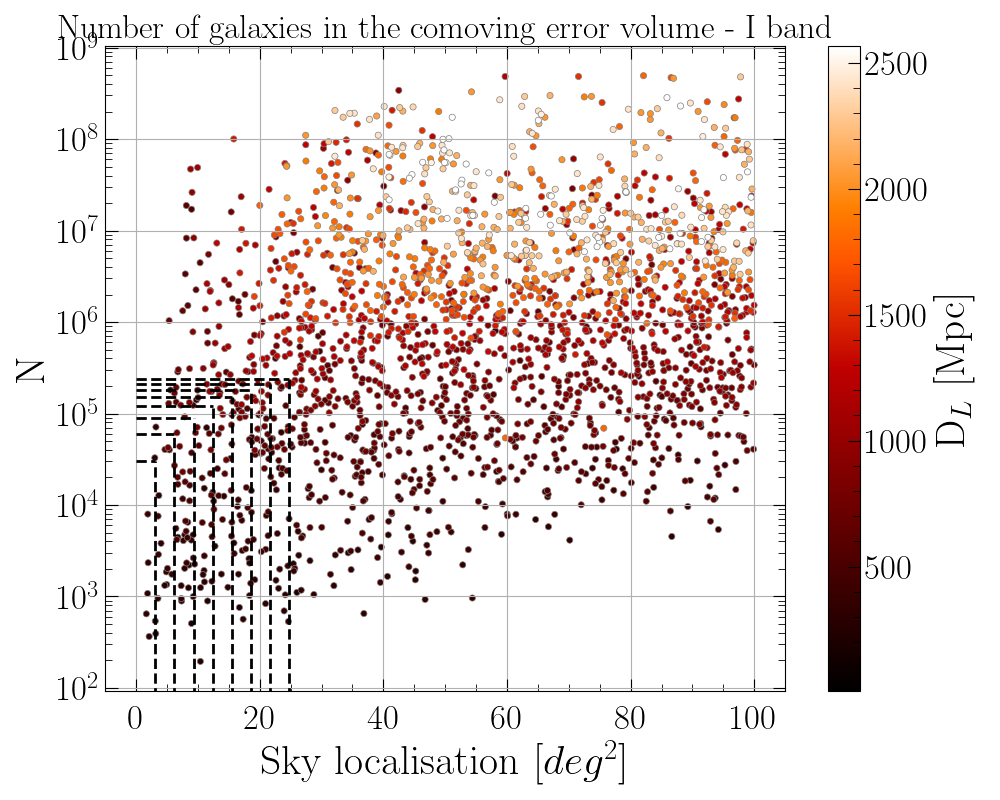}
    \caption{Number of galaxies in the comoving error volume of BNS at $z<0.45$ detected by ET alone, as a function of their sky localisation. Vertical dashed lines represent the sky area covered for increasing number of exposures. Horizontal dashed lines refer to the corresponding number of WST fibers available within the FoV considered. The color bar refers to the GW signal luminosity distance.}
    \label{ilbert_I_skyloc_ETonly}
\end{figure}

\begin{figure}
    \includegraphics[width = \linewidth]{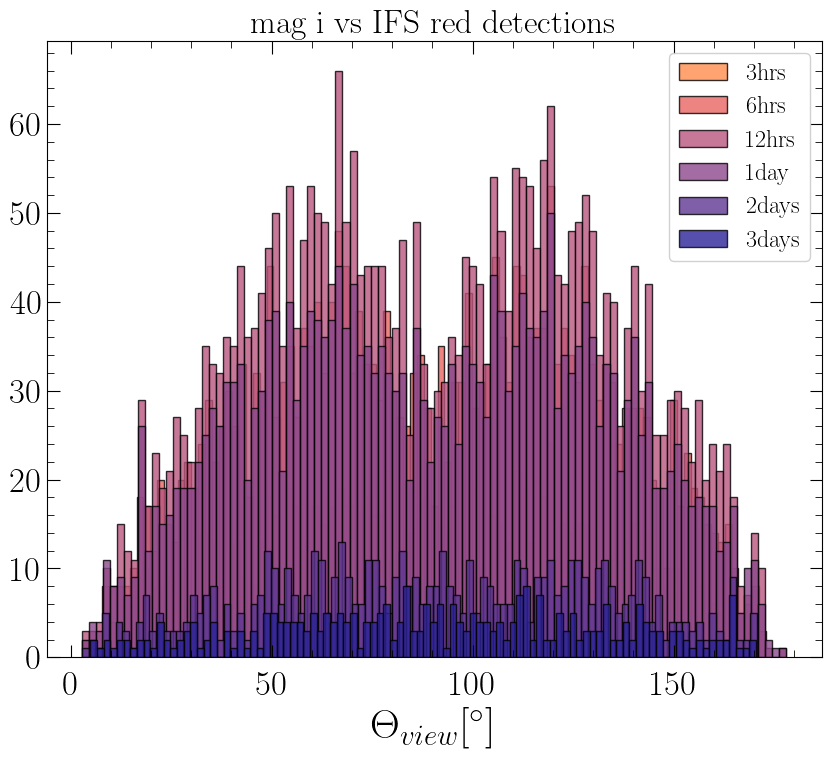}
    \includegraphics[width = \linewidth]{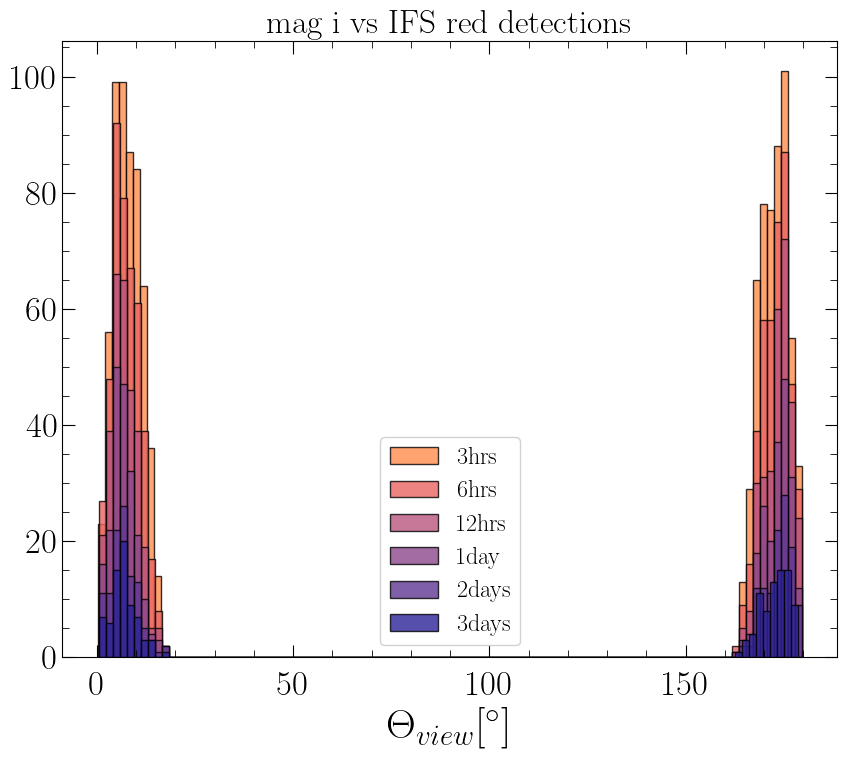}
    \caption{Viewing angle distribution of the population of events featuring an afterglow, that are detectable with WST IFS red arm, both when the afterglow outshines the KN (bottom) panel and viceversa (top). Different colors refer to different times post-merger. The GW interferometers network includes ET and CE.
    }
\end{figure}